\documentclass{article}

\usepackage{amsmath, amsthm, amssymb}
\usepackage{graphicx}
\usepackage{verbatim}
\usepackage{natbib}
\usepackage{caption}
\usepackage{subcaption}
\usepackage{fancyvrb}
\usepackage{enumerate}
\usepackage{relsize}
\usepackage{float}
\usepackage{xcolor}

\usepackage{hyperref}
\usepackage[margin=1.5in]{geometry}
\hypersetup{colorlinks,citecolor=blue,urlcolor=blue,linkcolor=blue}
\usepackage{booktabs}

\usepackage{placeins}

\usepackage{stefan_tex}

\newcommand\blfootnote[1]{%
  \begingroup
  \renewcommand\thefootnote{}\footnote{#1}%
  \addtocounter{footnote}{-1}%
  \endgroup
}

\newcommand{\rev}[1]{#1}

\theoremstyle{plain}
\newtheorem{prop}{Proposition}

\newtheorem{theo}[prop]{Theorem}

\theoremstyle{definition}

\newtheorem{assu}{Assumption}

\theoremstyle{remark}

\author{Chenyu Qiu \\ Uber Technologies \and
Xu Kuang \\ Stanford GSB \and 
Inessa Liskovich \\ Uber Technologies \and 
Ali Rauh \\ Uber Technologies \and
Stefan Wager \\ Stanford GSB}
\date{Draft version \ifcase\month\or
January\or February\or March\or April\or May\or June\or
July\or August\or September\or October\or November\or December\fi \ \number%
\year\ \  }
\title{What is the Long-Term Value of Reliability?}

\begin{document}

\maketitle

\begin{abstract}
We describe Chronos LTV, a system to measure the long-term impact of delays and other service defects on key business metrics. We use Markov decision processes to model customer interactions over time, and formalize our target estimand as the marginal policy effect with respect to moving the average delay rate. Given this setup, we show that we can identify long-term effects under a sequential unconfoundedness assumption where delays are as good as random given observed order characteristics; and can estimate these effects using a simple covariate-balancing algorithm.
\end{abstract}

\section{Introduction}

Running\blfootnote{\hspace{-5.3mm}Xu Kuang
contributed to this research while on leave at Uber. Stefan Wager contributed to this research as a paid consultant for Uber. In both cases, these contributions were not part of the author's Stanford University duties or responsibilities. Claude Code was used to assist with the implementation of the marketplace simulator described in Section \ref{sec:simulator}, as well as with the writing of the simulator specification in Appendix \ref{sec:simu_details}.}
a service operation requires navigating delicate trade-offs between costs and reliability. High reliability generally requires maintaining excess capacity and process flexibility, and maintaining such flexibility incurs costs \rev{that} then either directly or indirectly get passed on to consumers \citep{cachon2009matching}. For example, in the context of an on-demand food-delivery service, preventing the utilization level of couriers from getting too high improves reliability, but also means that couriers will spend more time idle between jobs and thus need to be paid more per job \citep{hall2023ride}. Service providers thus need to navigate a frontier of feasible system specifications that trace this cost-reliability trade-off, and face the challenge of choosing an operating point along this frontier that optimizes overall service quality and customer experience. Technological innovations, such as better forecasting or matching algorithms, can raise the Pareto frontier; however, the task of eventually choosing a point on the frontier remains.

A key challenge in balancing reliability with costs is that the costs of the flexibility needed to achieve reliability \rev{are} immediate and apparent, whereas its benefits take a long time to realize and are hard to measure. Consider, again, a food-delivery service that needs to choose whether to design its marketplace such as to achieve moderate (say 75\%) vs.~high (say 90\%) courier utilization. It's immediate that the moderate-utilization specification will incur roughly 20\% higher courier costs / delivery (assuming constant average earnings per hour for couriers). One can use tools from queuing theory (or short-term experiments) to understand how higher utilization will degrade reliability metrics, such as the frequency of delays \citep{hopp2001factory}. But it's much harder to measure how reliability translates into long-term consumer satisfaction and engagement.

This paper describes Chronos LTV, a system built by Uber to estimate the impact of delays on its food delivery platform using observational data. We model each customer (or ``eater'') as a Markovian agent with hidden states. The eaters occasionally place orders; and when they do, they update their hidden state (e.g., beliefs or preferences) based on their experience. These hidden states then inform future ordering behavior. We show how, under an assumption that delays are effectively random given observable order characteristics (such as time of day and basket size), we can identify the marginal policy effect (MPE) of shifting the delay rate from observational data. More specifically, we can use Chronos to make assessments such as: If we reduce the steady-state rate of delays by 1 percentage point, we expect to see long-term steady-state improvement of $X$ percentage points in key eater metrics.

The MPEs we estimate using Chronos are often referred to as policy gradients in the reinforcement learning literature \citep{sutton1998reinforcement}. Our approach to estimating these quantities builds on recent work by \citet*{johari2025estimation} who showed how policy gradients are identified from interaction-level-randomized data under mixing assumptions on the underlying Markov process; and by \citet{lai2026estimating} who showed that this identification remains valid under sequential unconfoundedness assumptions.

We present our model and method in Section \ref{sec:method}. Due to confidentiality reasons, we cannot share results from applying this method to Uber data publicly. In Section \ref{sec:simulator}, however, we describe a marketplace simulator that instantiates some qualitative properties of on-demand food-delivery marketplaces. In Section \ref{sec:exp1} we use our simulator to verify that Chronos can in fact correctly predict the outcomes of hypothetical long-term reliability experiments; while in Section \ref{sec:exp2} we demonstrate use of Chronos to reason about the long-term benefits of increasing courier supply.

\subsection{Related Work}

The conceptual gold standard for estimating causal effects is to run randomized experiments that implement proposed policies for long enough for effects to stabilize. Running long enough experiments, however, is often easier said than done. The required timelines may be misaligned with targets for innovation speed: For example, \citet{hohnhold2015focusing} report results from online advertising experiments where user behavior took multiple months to converge. In some contexts it may be possible to avoid long-horizon experiments by using surrogate endpoints \citep{athey2025surrogate}; however, surrogate methods generally require strong assumptions (including statistical surrogacy) that can be hard to verify and may not always hold.

Given the challenge---and often expense---of running long-term experiments, there is also considerable interest \rev{in methods} that can extrapolate from a sequence of short-term, random interventions to long-term effects. In the reinforcement learning literature, this task is typically referred to as off-policy evaluation. When we are in a Markovian system and all states that mediate long-term effects are observed, then statistically efficient methods for off-policy evaluation are available \citep{kallus2022efficiently,liao2022batch,mehrabi2024off}. In our setting, however, it is unlikely that all relevant states (e.g., eater beliefs or preferences) would be observed. And, unfortunately, generic methods for off-policy evaluation without fully-observed Markovian states---including doubly robust methods---generally have variance that scales exponentially with the length of the relevant decision horizon \citep{bang2005doubly,jiang2016doubly,robins2000marginal,thomas2016data}.

Our paper fits within a recent line of work that seeks to \rev{mitigate} the variance issues for generic off-policy evaluation by focusing on policy counterfactuals that are ``adjacent'' to the status quo. More specifically, this literature seeks to estimate the MPE of modifying continuous variables that determine the policy; in reinforcement learning terminology, one can frame the focus of this literature in terms of policy gradients \citep{sutton1998reinforcement}. The reason to focus on MPEs is that they admit simple and stable estimators using data collected from experiments with short-term interventions \citep{farias2022markovian,johari2025estimation,li2023experimenting}, as well as from similarly structured observational studies \citep{ghosh2025non,lai2026estimating}. Our paper contributes to this literature by demonstrating the promise and feasibility of MPE-focused methodology in the context of realistic deployment to an online marketplace setting.

\section{Chronos LTV}
\label{sec:method}

We observe data on $i = 1, \, 2, \, \ldots, \, n$ customers sampled at time points $t = 1, \, 2, \, \ldots, \, T$, and assume that the observed units are sampled IID from a population of customers. At each time point $t$, each customer has a latent state $U_{it} \in \uu$ characterizing, e.g., their beliefs, plans and satisfaction with past experience. Potentially informed by $U_{it}$ and/or exogenous covariates such as time of day, the customer then either chooses to place an order with observed attributes $X_{it} \in \xx$, or chooses to not place an order $X_{it} = \emptyset$. If the customer places an order, the order may or may not be delayed, $D_{it} \in \cb{0, \,1}$ (if $X_{it} = \emptyset$ then there is no delay and $D_{it} = 0$). Finally, we record a performance metric $Y_{it} \in \RR$ for customer $i$ in period $t$; overall, we prefer for this performance metric to be higher on average.

We define the ``value'' of the status-quo policy, i.e., the average reward rate under the status quo, as \smash{$V_0 = \mathbb{E}[T^{-1} \sum_{t = 1}^T Y_{it}]$}. The question Chronos seeks to answer is: How would the value change if delays became infinitesimally more (or less) common? More specifically, writing $V_{\varepsilon}$ as average reward rate we would obtain if the average delay rate were exogenously increased by $\varepsilon$, our goal is to estimate $\tau = \sqb{d/d\varepsilon \, V_\varepsilon}_{\varepsilon = 0}$. We view this quantity as an estimate of long-term value of delays, in that it doesn't ask how a specific delay impacts its proximal outcomes, but instead helps us understand how a change in the overall delay rate will impact our steady-state reward rate in the long run.

In order to rigorously define $V_\varepsilon$ and to identify $\tau$, we rely on the following assumptions throughout. The first assumption, substantively, requires that the latent $U_{it}$ mediate any long-term impacts of customer experience on customer delays. The second assumption, when paired with the first, is a sequential unconfoundedness assumption that enables identification of causal effects \citep{robins1986new}; a recent textbook discussion of assumptions of this type is given in \citet[Chapter 14]{wager2026causal}. The third assumption implies that effects of past delays on future outcomes eventually washes out; see \citet{hu2022switchback} and \citet{johari2025estimation} for use of similar assumptions in the context of dynamic causal inference.

\begin{assu}
\label{assu:markov}
The tuples $S_{it} = \p{U_{it}, \, X_{it}, \, D_{it}}$ across time $t = 1, \, \ldots, \, T$ form a Markov chain for each customer $i$. The rewards $Y_{it}$ are generated as a random function of $S_{it}$, with noise independent of everything else in the model.
\end{assu}

\begin{assu}
\label{assu:unconf}
Delays are effectively random after conditioning on the observed state,
\begin{equation}
 D_{it} \indep \cb{S_{i(t-1)}, \, U_{it}} \cond X_{it}.
\end{equation}
\end{assu}

\begin{assu}
\label{assu:mix}
The effects of past delays on future rewards decays over time: There are constants $C, \ \nu > 0$ such that, uniformly across all $x$, $t$ and $k \geq k_0$,
\begin{equation}
\begin{split}
&\abs{\EE{Y_{i(t+k)} \cond X_{it} = x, \, D_{it} = 1} - \EE{Y_{i(t+k)} \cond X_{it} = x, \, D_{it} = 0}} \leq C  e^{-k/\nu}.
\end{split}
\end{equation}
\end{assu}

Given these assumptions, we can use the mathematical formalism of Markov decision processes (MDPs) to define the value-counterfactuals $V_\varepsilon$ \citep{sutton1998reinforcement}. In this formalism, delays $D_{it}$ will play the analytic role typically played by actions. We emphasize that delays are not actions in the classical sense: We consider a setting where the system operator does not directly control delays and would prefer to never have delays, but recognizes that delays arise inevitably from congestion dynamics with stochastic supply and demand. In the mathematical formalism, however, delays correspond to actions in that they are ``inputs'' to our causal question, and we are interested in understanding how changes to the delay distribution would impact down-stream rewards.

Given Assumptions \ref{assu:markov} and \ref{assu:unconf}, we can factor the joint distribution $\cb{U_{it}, \, X_{it}, \, D_{it}, \, Y_{it}}$ as\footnote{To enable a compact form of the factorization, we define $U_{i(T+1)} = X_{i(T+1)} = \emptyset$.}
\begin{equation}
\label{eq:prob_fact}
\begin{split}
&\PP{U_{i1}, \, X_{i1}, \, D_{i1}, \, Y_{i1}, ,\, \ldots, \, U_{iT}, \, X_{iT}, \, D_{iT}, \, Y_{iT}} = \PP[1]{U_{i1}, \, X_{i1}} \\
&\quad\quad\quad\quad\quad\quad \times \prod_{t = 1}^T \PP[t]{D_{it} \cond X_{it}} \PP[t]{Y_{it} \cond S_{it}} \PP[t+1]{U_{i(t+1)}, \, X_{i(t+1)} \cond S_{it}}.
\end{split}
\end{equation}
The status-quo policy value $V_0$ is the expected average reward, $T^{-1} \sum_{t = 1}^T Y_{it}$, under the above distribution. We are interested in welfare counterfactuals that arise if we marginally modify delay rates whenever $X_{it}\neq \emptyset$ (i.e., whenever an order was placed and so delays are possible). For small values of $\varepsilon$, define
\begin{equation}
\label{eq:g-formula}
\begin{split}
&\mathbb{P}^{\varepsilon}\sqb{U_{i1}, \, X_{i1}, \, D_{i1}, \, Y_{i1}, ,\, \ldots, \, U_{iT}, \, X_{iT}, \, D_{iT}, \, Y_{iT}} = \PP[1]{U_{i1}, \, X_{i1}} \\
&\quad\quad\quad\quad\quad\quad \times \prod_{t = 1}^T \mathbb{P}_t^{\varepsilon}\sqb{D_{it} \cond X_{it}} \PP[t]{Y_{it} \cond S_{it}} \PP[t+1]{U_{i(t+1)}, \, X_{i(t+1)} \cond S_{it}}, \\
&\text{where } \mathbb{P}_t^{\varepsilon}\sqb{D_{it} = 1 \cond X_{it} = x} = \mathbb{P}_t\sqb{D_{it} = 1 \cond X_{it} = x} + \varepsilon.
\end{split}
\end{equation}
for all $x \neq \emptyset$.
In other words, we define delay-impacted distributions $\mathbb{P}^{\varepsilon}$ by adjusting the delay probabilities in \eqref{eq:prob_fact} while leaving all other conditional probabilities unchanged \citep{robins1986new}. Then, writing $\mathbb{E}^\varepsilon$ for expectations under the distribution $\mathbb{P}^{\varepsilon}$, we define\footnote{\rev{It is also possible to define other estimands that, e.g., only create delays in certain states or have variation in how susceptible states are to new delays. Our approach extends directly to such settings; however, we omit details for space and instead refer readers to \citet{lai2026estimating} for further discussion.}}
\begin{equation}
\label{eq:tau_def}
V_\varepsilon = \mathbb{E}^\varepsilon\sqb{\frac{1}{T} \sum_{t = 1}^T Y_{it}}, \ \ \ \ \tau = \sqb{\frac{d}{d\varepsilon} V_\varepsilon}_{\varepsilon = 0},
\end{equation}
and seek to estimate $\tau$. The key identifying result allowing us to do, given below, is a consequence of the policy-gradient theorem \citep{sutton1999policy}. The proof follows by synthesizing arguments made in \citet{johari2025estimation} and \citet{lai2026estimating}; a self-contained argument is given in Section \ref{sec:proof}.

\begin{theo}
\label{theo:pgt}
Under Assumptions \ref{assu:markov}--\ref{assu:mix}, suppose furthermore that the expectations and derivative given in \eqref{eq:tau_def} are finite and well defined, and that delay probabilities satisfy the following overlap condition
\begin{equation}
\label{eq:overlap}
\varepsilon \leq \PP[t]{D_{it} = 1 \cond X_{it} = x} \leq 1 - \varepsilon, \text{ for some $\varepsilon > 0$ and all $x \neq \emptyset$ and $t$.}
\end{equation}
For any horizon parameter $K \geq k_0$, define an oracle estimator\footnote{We refer to this estimator as an ``oracle'' because it depends on conditional delay probabilities $\PP[t]{D_{it} = 1 \cond X_{it}}$ which are typically unknown in practice---and would need to be estimated in order to form a feasible estimator.}
\begin{equation}
\label{eq:oracle_tau}
\begin{split}
\htau^*_K = \frac{1}{nT} \sum_{i = 1}^n \sum_{t = 1}^T 1\p{\cb{X_{it} \neq \emptyset}} \p{\frac{D_{it}}{\PP[t]{D_{it} = 1 \cond X_{it}}} - \frac{1 - D_{it}}{\PP[t]{D_{it} = 0 \cond X_{it}}}} \, \Gamma_{it}^K,
\end{split}
\end{equation}
with $\Gamma_{it}^K = \sum_{s = t}^{\p{t + K} \land T} Y_{is}$.
Then, we have $\abs{\EE{\htau^*_K} - \tau} \leq C / (1 - e^{-1/\nu}) \, e^{-K/\nu}$.
\end{theo}

Given the identification result in Theorem \ref{theo:pgt}, we see that consistent estimation of $\tau$ is possible under Assumptions \ref{assu:markov}--\ref{assu:mix} provided we can consistently estimate the conditional delay probabilities
\begin{equation}
\label{eq:pi}
\pi_t(x) = \PP[t]{D_{it} = 1 \cond X_{it} = x}.
\end{equation}
This object is closely analogous to a propensity score, i.e., a conditional treatment probability in an observational study under unconfoundedness, thus suggesting that we should be able to design feasible estimators motivated by propensity-score methods from causal inference \citep{imbens2004nonparametric}. In the section below, we outline one estimation strategy that adapts the covariate-balancing propensity score algorithm (CBPS) \citep{imai2014covariate,viviano2026dynamic,zhao2019covariate}. Other estimation strategies are, however, also possible; for example, \citet{lai2026estimating} provide results that can be used to build a doubly robust estimator for $\tau$ within the debiased machine learning framework \citep{chernozhukov2022locally}.

\subsection{A Covariate-Balancing Estimator}

The identification result above is non-parametric; however, to proceed with estimation, it is helpful to make further regularity assumptions. Here, we assume that delay probabilities are time-independent (unless time features are explicitly coded in $X_{it}$) and that, conditionally on an order being placed, delay probabilities satisfy a logistic functional form. We assume that $X_{it} \in \RR^d \cup \cb{\emptyset}$, and that there is a vector $\beta \in \RR^d$ such that\footnote{We also use an intercept in \eqref{eq:logistic}, but leave this implicit in the notation.}
\begin{equation}
\label{eq:logistic}
\pi_t(x) = \begin{cases}
0 & \text{ if } x = \emptyset, \\
1 \big/ \p{1 + e^{-x \cdot \beta}} & \text{ else.}
\end{cases}
\end{equation}
Given this specification, there are many possible ways to estimate the unknown parameter $\beta$. One familiar solution would be to fit $\beta$ via logistic regression. A better idea, however, is to fit $\beta$ using a method that explicitly enforces sample balance on the $X_{it}$ when plugged into \eqref{eq:oracle_tau}: Logistic regression and covariate-balancing methods both consistently recover $\beta$ in large samples, but covariate-balancing methods also eliminate imbalance in observables in finite samples, thus enabling more robust and accurate estimation of $\tau$ \citep{graham2012inverse}. Here, we use such a covariate-balancing method, implemented via the tailored loss function approach of \citet{zhao2019covariate}.

If the $\pi_t(x)$ are known, then inverse-weighting using these probabilities balances observables in expectation \citep{horvitz1952generalization}:
\begin{equation}
\label{eq:avg_balance}
 \EE{\sum_{\cb{i, t : X_{it} \neq \emptyset}} \frac{D_{it}}{\pi_t(X_{it})} X_{it}}
=  \EE{\sum_{\cb{i, t : X_{it} \neq \emptyset}} \frac{1 - D_{it}}{1 - \pi_t(X_{it})} X_{it}}
= \EE{\sum_{\cb{i, t : X_{it} \neq \emptyset}} X_{it}}.
\end{equation}
Covariate-balancing methods seek to estimate the parameter $\beta$ in \eqref{eq:logistic} by directly enforcing an empirical version of the balance condition \eqref{eq:avg_balance}. Specifically, we obtain two estimates \smash{$\hbeta_0$} and \smash{$\hbeta_1$} for the parameters in \eqref{eq:logistic} by solving the moment equations:
\begin{equation}
\begin{split}
&\frac{1}{nT} \sum_{i=1}^n \sum_{t=1}^T 1(\{X_{it} \neq \emptyset\}) \left(1 + e^{-X_{it}^\top \hat{\beta}_1}\right) D_{it} X_{it} = \frac{1}{nT} \sum_{i=1}^n \sum_{t=1}^T 1(\{X_{it} \neq \emptyset\}) X_{it}, \\
&\frac{1}{nT} \sum_{i=1}^n \sum_{t=1}^T 1(\{X_{it} \neq \emptyset\}) \left(1 + e^{X_{it}^\top \hat{\beta}_0}\right) (1 - D_{it}) X_{it} = \frac{1}{nT} \sum_{i=1}^n \sum_{t=1}^T 1(\{X_{it} \neq \emptyset\}) X_{it}.
\end{split}
\label{eq:CBPS_moment}
\end{equation}
The reason we needed to use a redundant parametrization (i.e., to produce two different estimates \smash{$\hbeta_0$} and \smash{$\hbeta_1$} for $\beta$) is that \eqref{eq:avg_balance} has \rev{$2d$ moments} whereas the model \eqref{eq:logistic} only has \rev{$d$ degrees of freedom}, so we need to introduce extra degrees of freedom for the system \eqref{eq:CBPS_moment} to have a solution in general.
Then, once the \smash{$\hat{\beta}_w$} are obtained by solving the above moment equations, the fitted probabilities are plugged directly into the propensity scores in \eqref{eq:oracle_tau}:
\begin{equation}
\label{eq:cbps_tau}
\begin{split}
\htau_K = \frac{1}{nT} \sum_{i = 1}^n \sum_{t = 1}^T 1\p{\cb{X_{it} \neq \emptyset}} \p{D_{it}\p{1 + e^{-X_{it}^\top \hat{\beta}_1}}  - \p{1 - D_{it}}\p{1 + e^{X_{it}^\top \hat{\beta}_0}}} \, \Gamma_{it}^K,
\end{split}
\end{equation}
Confidence intervals are obtained via a unit-clustered bootstrap \citep{efron1994introduction}: We draw bootstrap samples of units $i$ sampled independently with replacement; re-run the whole procedure, including solving \eqref{eq:CBPS_moment} and forming \eqref{eq:cbps_tau}, on each bootstrap sample; and estimate the standard error of $\htau_K$ via the standard deviation of the bootstrapped estimates.

The moment conditions in \eqref{eq:CBPS_moment} are non-linear in $\beta_w$, and so solving them numerically may at first appear daunting. The key to an efficient algorithm here is to notice that these moments can be re-interpreted as KKT conditions for an empirical minimization problem with a convex loss function: Whenever a finite-valued solution to \eqref{eq:CBPS_moment} exists, it can equivalently be characterized as
\begin{equation}
\begin{split}
&\hbeta_0 = \argmin_\beta\cb{\sum_{\cb{i, t : X_{it} \neq \emptyset}} {(1 - D_{it}) e^{X_{it}^\top\beta_0} - D_{it} X_{it}^\top\beta_0 }}, \\
&\hbeta_1 = \argmin_\beta\cb{\sum_{\cb{i, t : X_{it} \neq \emptyset}} {D_{it} e^{-X_{it}^\top\beta_1} + (1 - D_{it}) X_{it}^\top\beta_1 }}.
\end{split}
\end{equation}
This immediately implies that if a solution to \eqref{eq:CBPS_moment} exists then this solution is unique, and can be recovered via a convex loss minimization algorithm such as Newton descent. We refer to \citet[Chapter 7]{wager2026causal} for further details, including a discussion of double-robustness properties of covariate-balancing methods, and of conditions under which solutions to \eqref{eq:CBPS_moment} exist with high probability.

\section{A Simulator for On-Demand Food Delivery}
\label{sec:simulator}

We now describe a marketplace simulator we will use to demonstrate how Chronos LTV can recover long-term equilibrium impact of policy changes that improves delivery reliability, even when market participants have unobserved dynamics. We emphasize that this simulator is not intended to reflect actual production systems, policies or operating parameters of any real-world marketplace. Rather, it is designed to illustrate stylized feedback loops that arise in a multi-sided delivery marketplace, and assess the ability of Chronos LTV to capture this type of system dynamics.

The simulator consists of several interacting pieces: eaters, couriers, marketplace matching/dispatch, and delivery outcomes. Together they reproduce a marketplace where delays are not fully random, but rather vary with endogenous market conditions and user evolution. Failing to account for the endogeneity and dynamics will thus generally result \rev{in} estimators that miss the true policy effects by an order of magnitude.

The overall structure of the simulator is as follows. On any given day, each eater may choose to place an order, \rev{with ordering decisions governed} by a variety of latent factors described further below. Their orders will then be matched and dispatched by the marketplace and delivered by a courier. \rev{Stochastic variation in supply, matching, food preparation, travel, etc., can lead to delays in delivery.} Experiencing delays will then reduce the eater's propensity to place orders in upcoming days \rev{and their long-term engagement}. The strength of the response to delays varies by eater-specific delay sensitivity and slowly decays over time. Multiple delays can also compound: For example, while one delay may reduce an eater's order propensity by 30\%, a second consecutive delay may further drop it to below 50\%. \rev{While delays under our model are far from being uniformly random, they satisfy sequential unconfoundedness (Assumption \ref{assu:unconf}) by construction. This is because, conditionally on the placed order, delays only depend on the state of the marketplace and do not depend on latent eater attributes.}

When reporting experiment results, we always run the whole simulator over at least 250 ``days''. We use the first 100 days as burn-in, and then conduct our analysis on data collected after the burn-in period is over. Appendix Figure \ref{fig:lifecycle_composition} illustrates convergence of our simulator during the burn-in phase.

\paragraph{The Matching Process} The orders described above are serviced by a pool of couriers, who are paired with orders through a matching process. The marketplace runs a 5-minute slot loop each day, drawing courier supply and matching orders based on ``First-In-First-Out'' (FIFO) queue and a courier acceptance model. Each courier in the pool is shown the \rev{top-4} queued orders and decides which one to take according to an acceptance probability that depends on individual preference and each order’s attributes such as distance and tip. Matched couriers will start the delivery and be removed from the queue. Unmatched orders remain in the queue for matching in future time slots until they have waited more than 80 minutes, after which they are marked as unfulfilled.

The prevalence of delays is largely governed by whether sufficient supply exists to meet demand at any given time. In our simulator, we model couriers as joining the marketplace according to a stochastic arrival process with varying rates based on the day of the week and the hour of the day. We also incorporate daily supply shocks that modulate the arrival rate of couriers; these shocks represent weather, local events, or driver-side disruptions that are unpredictable beforehand.

\begin{figure}
    \centering
    \includegraphics[width=\linewidth]{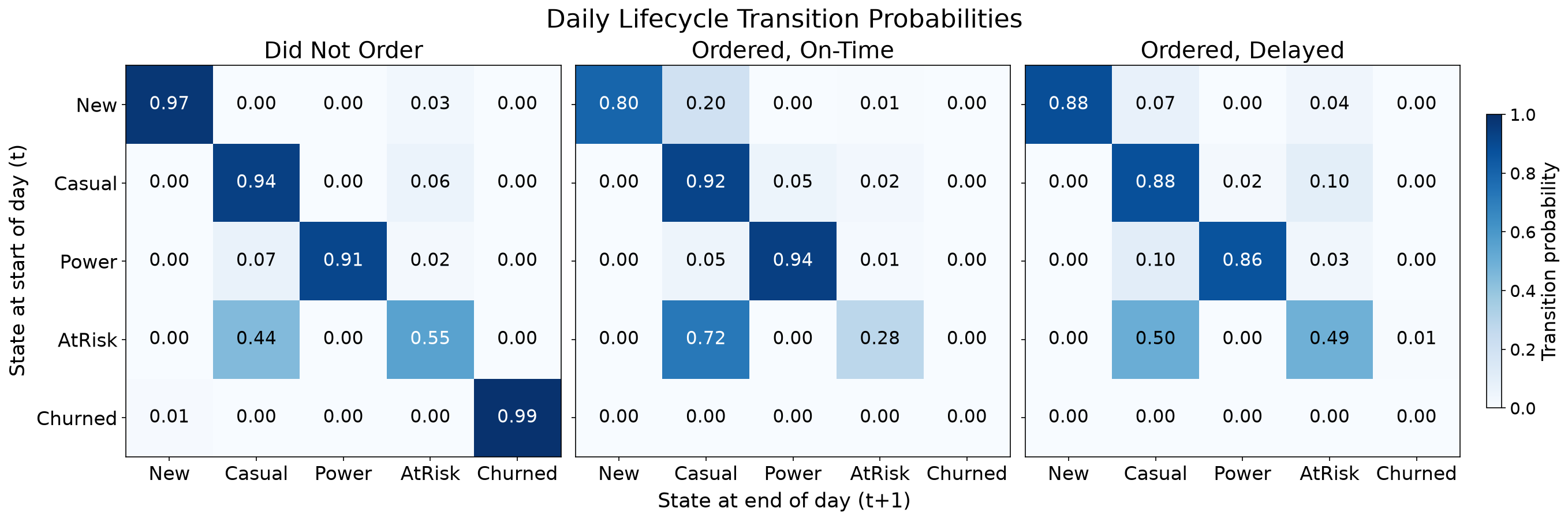}
    \caption{Daily lifecycle transition probabilities, by whether the eater did not order (left), ordered and was on-time (middle), or ordered and was delayed (right). Each cell shows row-normalized conditional transition probability.}
    \label{fig:lifecycle}
\end{figure}

\paragraph{Eater evolution} Eaters' choices in our simulator are governed by 7 latent attributes: baseline order rate, delay sensitivity, typical order value, tip generosity, explorativeness, and \rev{preferences over delivery distance and restaurant preparation time}. We simulate correlations between these attributes, e.g., eaters who order frequently are less delay-sensitive, delay-sensitive eaters tend to place orders with shorter delivery distances, eaters who place large orders tend to tip more. In addition to these persistent attributes, we model eaters as moving through 5 lifecycle states (see Appendix for further discussion): New, Casual, Power, AtRisk and Churned. As illustrated in Figure \ref{fig:lifecycle}, transitions between these lifecycle states are governed by past orders, past spending, recent activity and accumulated delay experiences.

\begin{figure}
    \centering
    \includegraphics[width=1\linewidth]{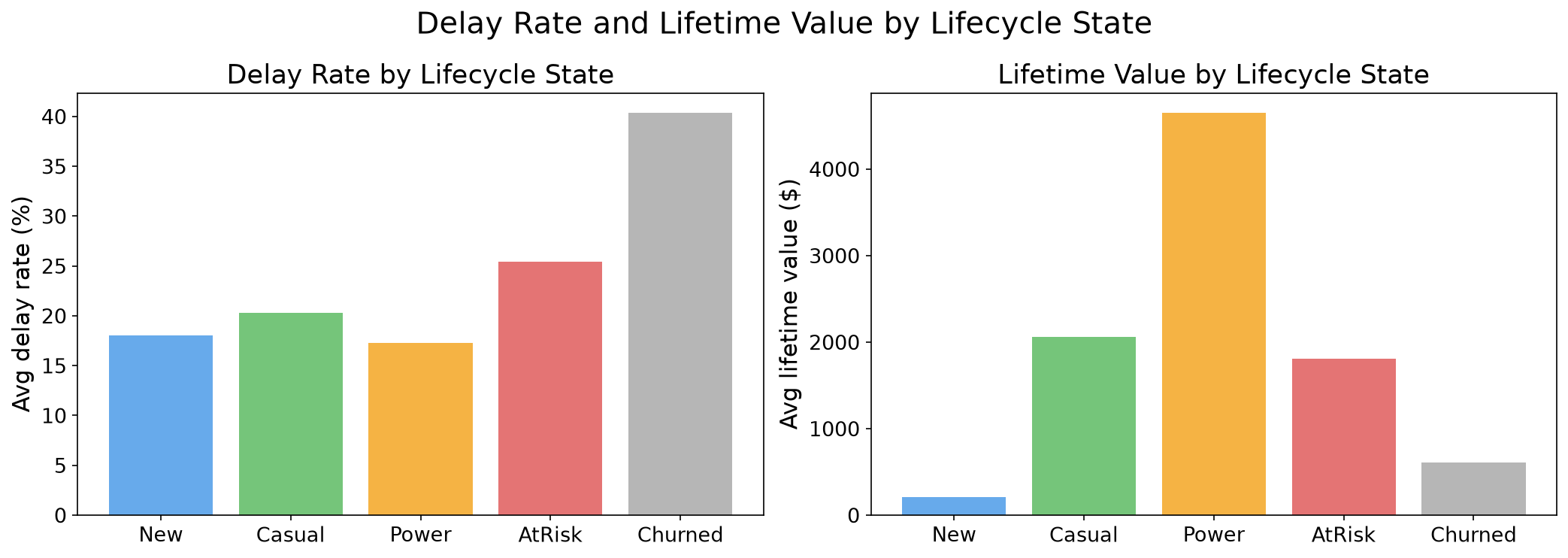}
    \caption{\rev{Eater-level outcomes grouped by lifecycle state. Lifecycle state is measured at day 250. {\it Left:} Average eater delay rate over matched orders during simulation days 150–249, among eaters with at least one such order. {\it Right:} Average cumulative spending through day 250, including all eaters in each state.}}
    \label{fig:confounding}
\end{figure}

We emphasize that although the analyst is able to see order metrics (such as order size and order frequency), the persistent attributes and lifecycle states described above cannot be directly observed. Furthermore, these unobservables are generally correlated with delays over time. For example, as shown in Figure \ref{fig:confounding}, eaters who end the simulation (at 250 days) in a ``Churned'' state will have much higher average delay rates.\footnote{Given knowledge of the simulator design, it's clear this is an example of reverse causality: There are many eaters who churned because they experienced too many delays.} However, our key unconfoundedness Assumption \ref{assu:unconf} still holds: Conditionally on observed order characteristics, delays are caused by market fluctuations (and so are independent of private eater attributes).

\begin{figure}
    \centering
    \includegraphics[width=1\linewidth]{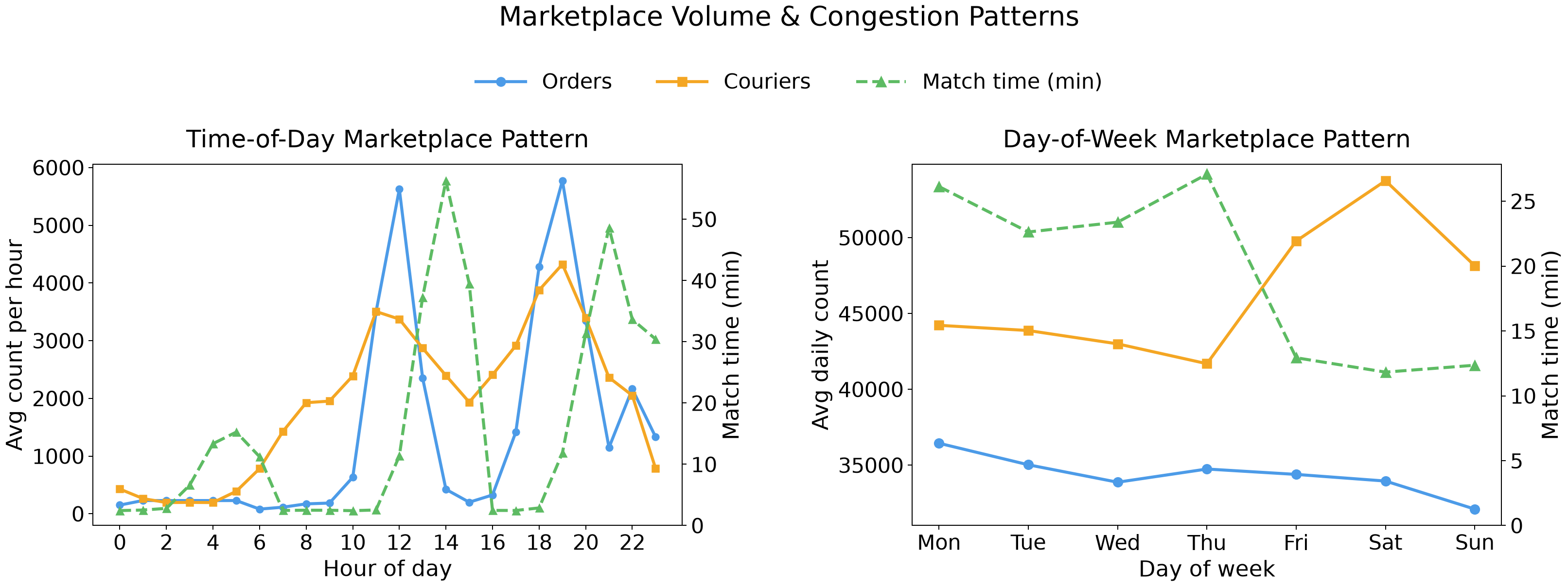}
    \caption{Marketplace volume and congestion patterns averaged over sims days. {\it Left:} time-of-day pattern showing orders per hour (blue), couriers per hour (orange), and the mean match time in minutes (green). {\it Right:} day-of-week pattern with daily orders (blue), daily couriers (orange), and average daily match time (green).}
    \label{fig:delays}
\end{figure}

\paragraph{Delay realization} Each time an eater places an order, they are quoted an expected time to delivery (ETD) based on expected food preparation time, matching time, and courier travel time. An order is deemed delayed if the actual delivery time exceeds ETD by 20 minutes or more.
Figure \ref{fig:delays} illustrates market outcomes that emerge from our simulator. We observe several qualitative findings that mimic typical business patterns for a delivery platform. In the left panel, orders follow a lunch (12pm) and dinner (7–8pm) cycle. Courier supply tracks demand; however, following a spike in demand, subsequent matching times get longer (because couriers are busy). These longer matching times happen after demand surges (instead of at their peak) because the matching queue takes time to build up---but then eventually creates spillovers from early orders to later orders. In the right panel, courier supply dips mid-week and rises during weekend while demand stays relatively flat, creating a supply-demand mismatch that makes average matching time fluctuate through the week.

\section{The Benefits of Faster Dispatch}
\label{sec:exp1}

Our first experiment considers evaluation of an intervention where we reduce eater delay exposure through an improved dispatch algorithm. Under the baseline condition, orders enter the matching queue only after food preparation is complete, so delivery is strictly sequential---and long matching times then directly increase delivery delays. We then evaluate an intervention that uses a proactive dispatch algorithm that allows couriers to be dispatched up to 5 minutes before food preparation finishes. It is immediately apparent (e.g., by running a small short-term experiment) that such proactive matching helps reduce delays in our simulator: The intervention reduces the delay rate roughly by 4 percentage points. The question of interest is how this translates into improved long-term value.

A straight-forward way to measure the long-term benefits of faster dispatch would be to run a long-term eater experiment: We could randomize a pool of users into treatment and control and then, over the course of a long experiment window (multiple months) give treated (but not control) users access to proactive dispatch. This kind of experiment, however, is expensive and organizationally challenging (the long timescales make fast decision-making difficult); and we would prefer being able to switch the control users to the new (better) matching algorithm faster.

\begin{figure}[t]
    \centering
    \small
    \includegraphics[width=1\linewidth]{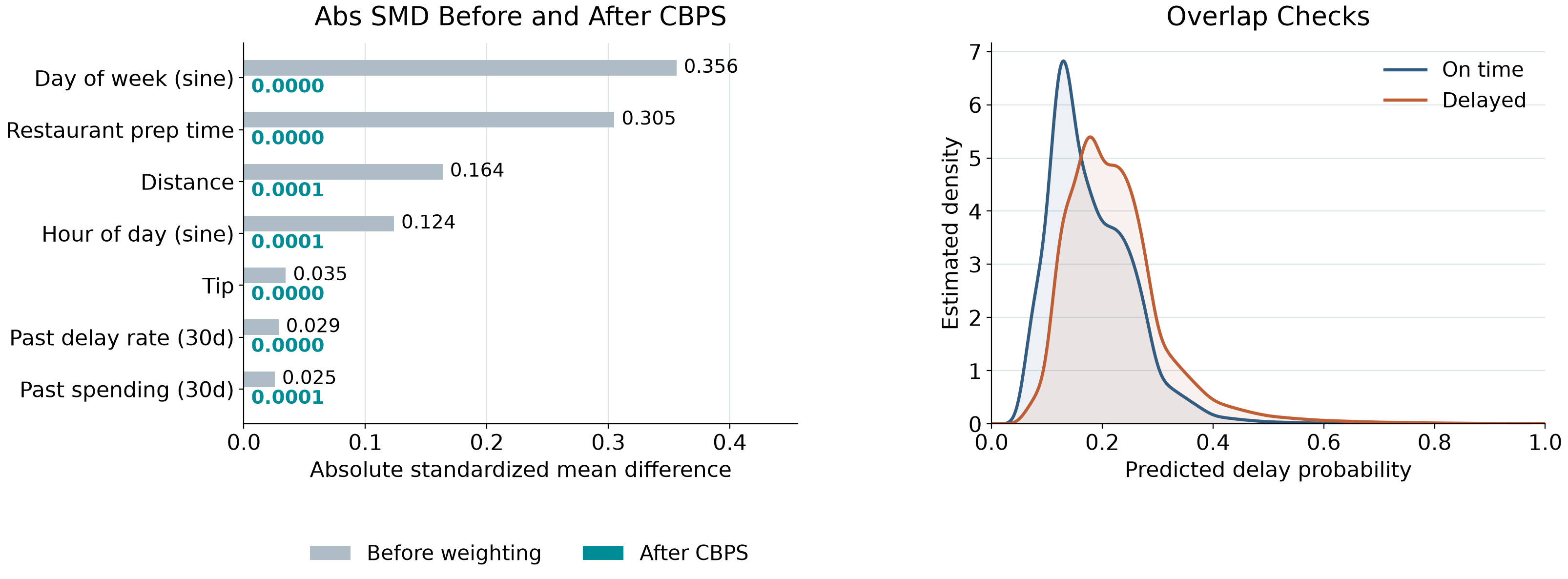}
    \caption{\rev{The left panel shows absolute standardized mean differences on representative covariates between delayed and on-time orders before and after CBPS weighting. The right panel shows the distribution of predicted delay probabilities by actual delay status.}}
    \label{fig:diagnostics}
\end{figure}

Here, we evaluate the ability of Chronos LTV---and baselines---to predict the outcome of this long-term user experiment from observational data. Our experiment proceeds in two phases. We first initialize a market with 100K eaters and run the simulator for 250 days under baseline dynamics. This observation period allows eater behaviors and lifecycle transitions to stabilize. During the data collection, we log comprehensive per-order and eater-level features as listed in Appendix Table \ref{tab:ltv_features}. Based on this data, we compute three estimators:
\begin{itemize}
    \item \textbf{Naive estimator:} This is not a Chronos estimate, and only computes the raw difference in future outcomes between delayed and non-delayed orders.
    \item \textbf{MLP-IPW estimator:} This follows Chronos framework and applies the same IPW formula \eqref{eq:oracle_tau}, but it estimates delay propensities using a multi-layer perceptron (MLP) trained on the observed features. The objective of the MLP is to maximize prediction accuracy rather than explicitly enforce covariate balance.
    \item \textbf{CBPS-IPW estimator:} This follows Chronos framework and replaces the standard propensity model with covariate balancing propensity scores \eqref{eq:cbps_tau}, where parameters are estimated to directly solve covariate balance moment conditions across all features while simultaneously fitting the exposure mechanism.
\end{itemize}
\rev{We first examine covariate balance and propensity score overlap to confirm that we make the exposed and unexposed groups comparable with CBPS reweighting. As shown in Figure \ref{fig:diagnostics}, delayed and on-time orders differ substantially on the covariates that predict delays such as day of week, restaurant prep time and delivery distance. CBPS directly targets the imbalance and almost fully removes these mean differences after reweighting. We also compare the distribution of estimated delay probabilities across delay and on-time orders to confirm overlap as assumed in \eqref{eq:overlap}.}

Each of these estimators has a horizon parameter $K$; we run all these estimators for $K = 56$ and 70 days. Each estimator represents the marginal effect of one percentage point change in the delay rate. To predict the LTV impact of faster dispatch, we therefore scale each estimate by the expected short-term delay effect of \(-4.15\) percentage points, assuming this effect has been learned from a prior pilot experiment.

After forming the Chronos estimates (i.e., after reaching 250 days), we next run the simulator for another 120 days during which time we simulate the gold-standard, long-term user-randomized experiment. At this point, 50\% of eaters are randomized to the treatment group \rev{and} get proactive dispatch as we described before, whereas the other 50\% are randomized to control and continue under the baseline dispatch policy. The question is then, to what extent is Chronos---using observational data collected at 250 days---able to predict the readout from this long-term experiment?

Table \ref{tab:recovery_k70} presents how predictions from different estimators compare to the observed steady state impact in last 4 weeks of the experiment. The naive
  estimator which ignores treatment selection entirely produces a massively
  inflated effect. IPW with MLP-estimated propensities removes most of that bias but is still 3x too large. The bootstrap CI of \rev{[+10.2\%, +23.6\%]} is also too wide and  still does not cover the truth. IPW with CBPS, by contrast, correctly predicts \rev{92--96\%} of the experiment-observed long-term effect, and both predictions (K=56d and K=70d) land
  within the range [+4.99, +6.65], i.e., the confidence interval obtained from the long-term user experiment we use as ground truth. Its bootstrap CI is also much smaller than the other two. This illustrates the robustness of CBPS to finite-sample imbalance and to potential misspecification of the propensity model. 

\begin{table}[t]
\centering
\small
\begin{tabular}{ccccccc}
\toprule
\textbf{$K$} & \textbf{Estimator} & $\boldsymbol{\beta}$ Mean & \textbf{$\beta$ 95\% CI} & \textbf{$\Delta$LTV\%} & \textbf{$\Delta$LTV 95\% CI} & \textbf{Calibration} \\
\midrule

56 & Naive
   & $-14.76$ & $[-14.87,-14.64]$
   & $+61.24\%$ & $[+60.76\%,+61.70\%]$ & $10.52\times$ \\
56 & MLP-IPW
   & $-4.11$ & $[-5.69,-2.45]$
   & $+17.04\%$ & $[+10.15\%,+23.60\%]$ & $2.93\times$ \\
56 & CBPS-IPW
   & $-1.29$ & $[-1.32,-1.25]$
   & $+5.34\%$ & $[+5.17\%,+5.47\%]$ & $0.92\times$ \\
\midrule
70 & Naive
   & $-18.35$ & $[-18.51,-18.19]$
   & $+76.13\%$ & $[+75.48\%,+76.78\%]$ & $13.08\times$ \\
70 & MLP-IPW
   & $-4.80$ & $[-7.04,-2.30]$
   & $+19.93\%$ & $[+9.55\%,+29.19\%]$ & $3.43\times$ \\
70 & CBPS-IPW
   & $-1.35$ & $[-1.41,-1.29]$
   & $+5.60\%$ & $[+5.37\%,+5.86\%]$ & $0.96\times$ \\
\bottomrule
\end{tabular}
\caption{Prediction of long-term impact under different estimators with bootstrapped confidence intervals. Each of these estimates starts by estimating a coefficient $\beta$ for the long-term value of delays; the final LTV estimate is then $-4.15 \times \beta$, where $-4.15\%$ is the pre-computed effect of the intervention on the delay rate. The ground-truth treatment-effect estimate, obtained by averaging outcomes in the long-term user experiment over weeks 14--17, is +5.82\% ([+4.99, +6.65]).}
\label{tab:recovery_k70}
\end{table}

\begin{figure}[t]
    \centering
    \includegraphics[width=1\linewidth]{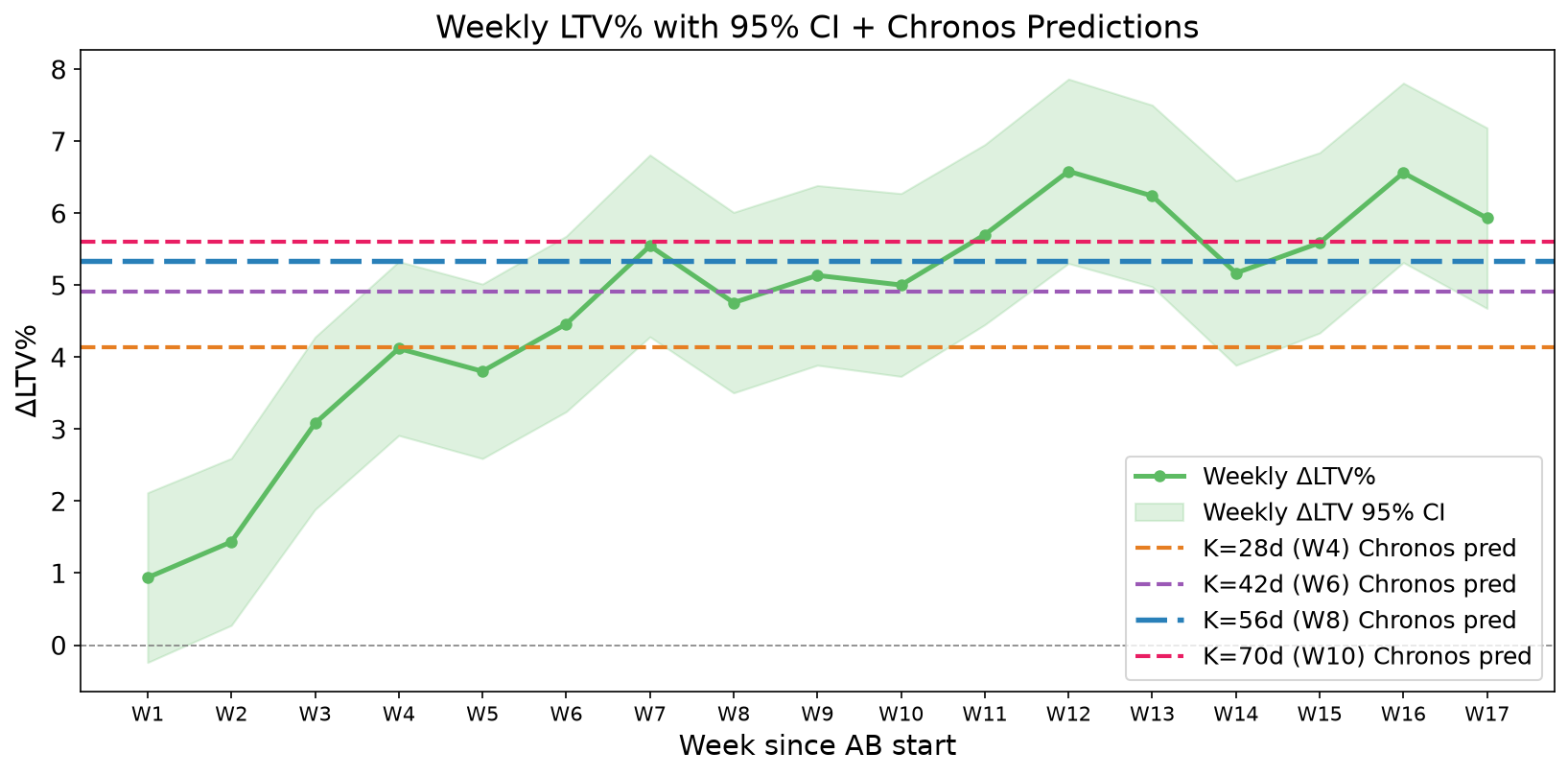}
    \caption{Validation of the Chronos estimator against the AB experiment ground truth. The green line is the weekly observed treatment effect on eater value. The shaded band is the per-week 95\% confidence interval. The dashed horizontal lines are Chronos predictions of the long-run effect with different forward-looking horizons (K days). Early K-day estimators line up well with the XP-observed impact at the matching week, with K=56d and K=70d converging to the full steady state impact.}
    \label{fig:lt_pred}
\end{figure}

Digging deeper, Figure \ref{fig:lt_pred} illustrates how the weekly treatment effect estimates on eater value in the long-term experiment evolves over its 17-week experimental window. We immediately see that the treatment effect ramps slowly over time: It is only +0.9\% in week 1, climbs through +4\% by week 4, and does not stabilize until week 12--14, where it plateaus around +5.5–6\%. This lag reflects behavioral adjustments over time: Treated and control eaters see the same delay-rate gap from day one, but lifecycle transitions and delay-memory decay both take weeks to propagate into ordering frequency and retention/churn, so the observable gap compounds over time toward its new equilibrium.

Figure \ref{fig:lt_pred} also shows Chronos predictions (obtained using the CBPS method) generated with horizons $K \in \{28, 42, 56, 70\}$. We see that not only does Chronos with $K = 56$ or 70 correctly anticipate the eventual full-experiment finding, but in fact running Chronos with different values of $K$ qualitatively predicts the evolution of the experiment readout over time. Chronos with $K = 28$ roughly predicts what we will measure at 4 weeks, using $K = 42$ predicts readouts at 6 weeks, etc.

\begin{figure}[t]
    \centering
    \small
    \includegraphics[width=0.7\linewidth]{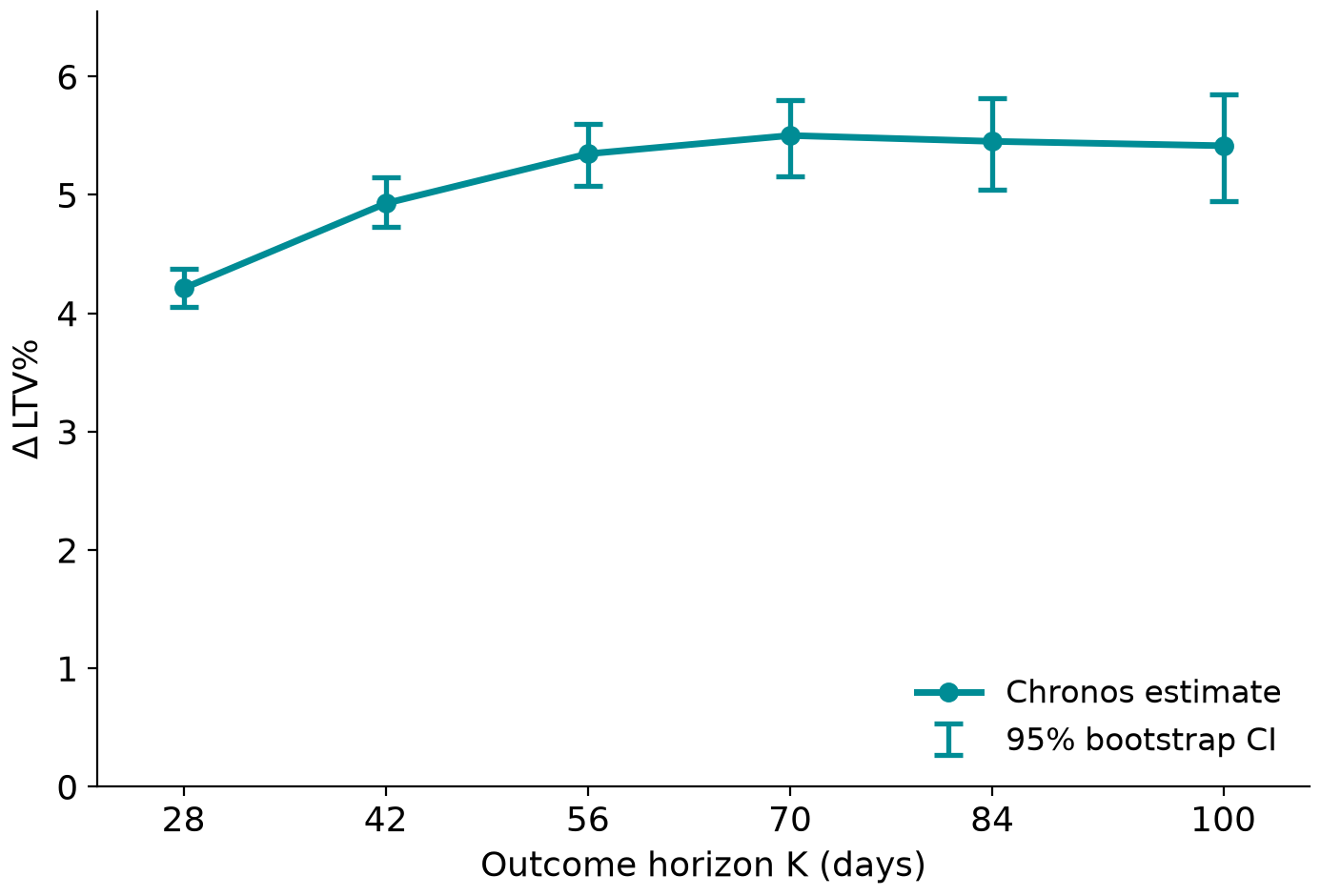}
    \caption{\rev{This figure illustrates how to choose the Chronos outcome horizon \(K\) in practice. Holding the cohort and observational window fixed, we can extend the forward-looking horizon and track the trend of the estimates until it stabilizes. In this simulation, the estimates converge around \(K=56\) or \(K=70\) days.}}
    \label{fig:k_converge}
\end{figure}

\rev{The horizon $K$ is a tuning parameter that requires careful choice by the analyst. We emphasize, however, that because $K$ is a a parameter chosen during the data analysis (rather than data collection) phase, it is possible to sweep across a range of plausible values of $K$ as in Figure \ref{fig:k_converge}, and to pick $K$ large enough for the response curve to converge. A more formal version of this approach (beyond the scope of this work) could proceed using Lepski's method \citep{lepskii1992asymptotically,su2020adaptive}.}

We end this section with a caveat. By reducing the number of delays, the improved dispatching algorithm here may increase long-term demand---thus creating market spillovers. Here, neither Chronos nor the long-term experiment we use as a gold standard capture such spillovers. Instead, the long-term experiment captures what \citet{li2022network} refer to as the long-term direct effect of the policy change; and we are thus effectively evaluating the ability of Chronos to also estimate this quantity. Getting full market-level estimates of the policy change may also require use of a spillover correction; however, the design of such corrections is typically orthogonal to how we estimate long-term direct effects. In the next section, we illustrate how Chronos estimates can be actually used to recover the full long-run impact of a market-wide rollout, by combining a market-level experimental design to capture the short-term equilibrium impact on delay reduction.

\section{The Value of Increasing Supply}
\label{sec:exp2}

We next illustrate how Chronos can be used for product evaluation and policy decision making in practice: If we persistently increase courier supply, how much does long-term value grow?
A direct experiment that captures the full long-run policy effect is difficult. For example, one design would be to run a long-horizon market-level experiment, such as randomizing cities into persistent supply increases. It would capture both the short-term market equilibrium response and the long-term eater lifecycle response from the increased supply. In practice, however, city-level experiments are expensive, slow, and often underpowered because the number of independent markets is limited.

Instead, with Chronos we can combine two powerful tools and two sources of evidence for this policy evaluation. We first used a switchback (SB) experiment to capture the immediate, short-term market-level effect of additional courier supply on delivery delays, including short-term spillovers through shared courier supply and matching efficiency \citep{bojinov2023design,hu2022switchback}. We then use Chronos to estimate how changes in delivery delay rates translate into future eater value using observational order-level data. Our final estimate of the LTV of persistent supply change is then a product of these two components:\footnote{\rev{Our estimation strategy implicitly assumes that the delay-effects of increasing supply are uniformly spread across states, thus connecting to the Chronos estimand \eqref{eq:tau_def}. In our current study, we did not attempt to correct for potential heterogeneity in the effect of the policy on delays as, in practice, we believe it would be hard to accurately measure this type of heterogeneity in a switchback. In settings where strong heterogeneity in delay response is suspected, alternative estimation strategies may need to be developed.}}
\begin{equation}
\label{eq:prod_est}
\widehat{\Delta LTV}^{\,\text{long-run}}_{\text{SB} \times \text{Chronos}}
=
\widehat{\Delta \text{Delay}}^{\,\text{SB}}
\times
\widehat{\text{LTV response to delay.}}^{\,\text{Chronos}} 
\end{equation}
In this study we estimate each component separately and test whether the combined \(\text{SB} \times \text{Chronos}\) approach can recover the total policy effect.

With our simulator, we can run a switchback for the policy change and also construct the ground truth directly. In each simulation run we initialize 100{,}000 eaters and then proceed through the following stages:
\begin{enumerate}
    \item \textbf{Warm-up + observation phase.} The market runs under the baseline policy for 100 days, allowing eater behavior, lifecycle composition, and matching dynamics to stabilize. The market continues under baseline conditions for another 150 days. During this period, we record same order-level and eater-level data as before, and use the data to estimate the Chronos LTV model (for K=56 days).

    \item \textbf{Forked evaluation phase.} At day 250, we fix the full market state, including each eater's order history, lifecycle state, and the current matching environment. From this common snapshot, we replicate the market into three indentically initialized and non-interacting segments that evolve in parallel (so in particular there are no spillovers between the segments):
    \begin{itemize}
        \item \textbf{Switchback experiment.} For 28 days, courier supply alternates every 3 hours between baseline supply and \(+25\%\) supply. We use stratified randomization to make sure treatment and control land on the same mix of busy and quiet times of week (day-of-week \(\times\) block-of-day). 

        \item \textbf{Always-control condition.} Baseline courier supply continues for 120 days. This represents the no-policy-change world. 

        \item \textbf{Always-treated condition.} Courier supply is increased by \(+25\%\) every day for 120 days. This represents the persistent policy rollout world.
    \end{itemize}
\end{enumerate}
We emphasize that, in the forked evaluation phase, both counterfactual deployment forks and the switchback fork all have the same demand and supply shock realizations, so their macro-level uncertainties match.

In the simulated switchback experiment, we estimate the short-term equilibrium effect of additional courier supply on delivery delays and on total eater values. We use a simple difference-in-means estimator with burn-in periods dropped \citep{bojinov2023design,hu2022switchback}:
\[
\hat{\tau}_{\mathrm{DM}}
=
\frac{1}{|\mathcal{B}_1|}
\sum_{b \in \mathcal{B}_1} \bar{Y}_b
-
\frac{1}{|\mathcal{B}_0|}
\sum_{b \in \mathcal{B}_0} \bar{Y}_b,
\]
where \(\bar{Y}_b\) denotes the average outcome in block \(b\), and \(\mathcal{B}_1\) and \(\mathcal{B}_0\) denote the treated and control blocks, respectively. We drop the first hour of each block as a burn-in period so that orders placed immediately after a potential treatment switch, which may still be matched against couriers under the previous regime, do not contaminate the treatment-control contrast. We construct a 95\% confidence interval using a day-clustered jackknife, given that all blocks within a day share supply and demand shocks.

We define the ground-truth long-run policy effect by comparing the two long-run counterfactual worlds we simulate: One in which courier supply remains at baseline, and another in which courier supply is persistently increased. The difference in long-run value between these two worlds gives the ground-truth total policy effect, including both the short-term equilibrium delay reduction and the induced long-term eater lifecycle response. Specifically,  we consider the difference in average eater values over the final 21 days of the 120-day evaluation window in the always treated condition vs.~the always control condition as the ground truth of total policy effect.\footnote{We also consider the difference in average delay rate in the first 28 days of the evaluation window as the ground truth for the short-term delay effect.}

We first test the performance of the estimators under a single, fixed market realization. We hold the exogenous environment constant---the same supply and demand shock paths and the same courier-arrival realization are reused in every replication---and re-simulate only the eater paths from run to run. So the only source of variation across runs is eater-path randomization and the only source of variation within a run is the treatment policy. 

\begin{figure}[t]
    \centering
    \includegraphics[width=1\linewidth]{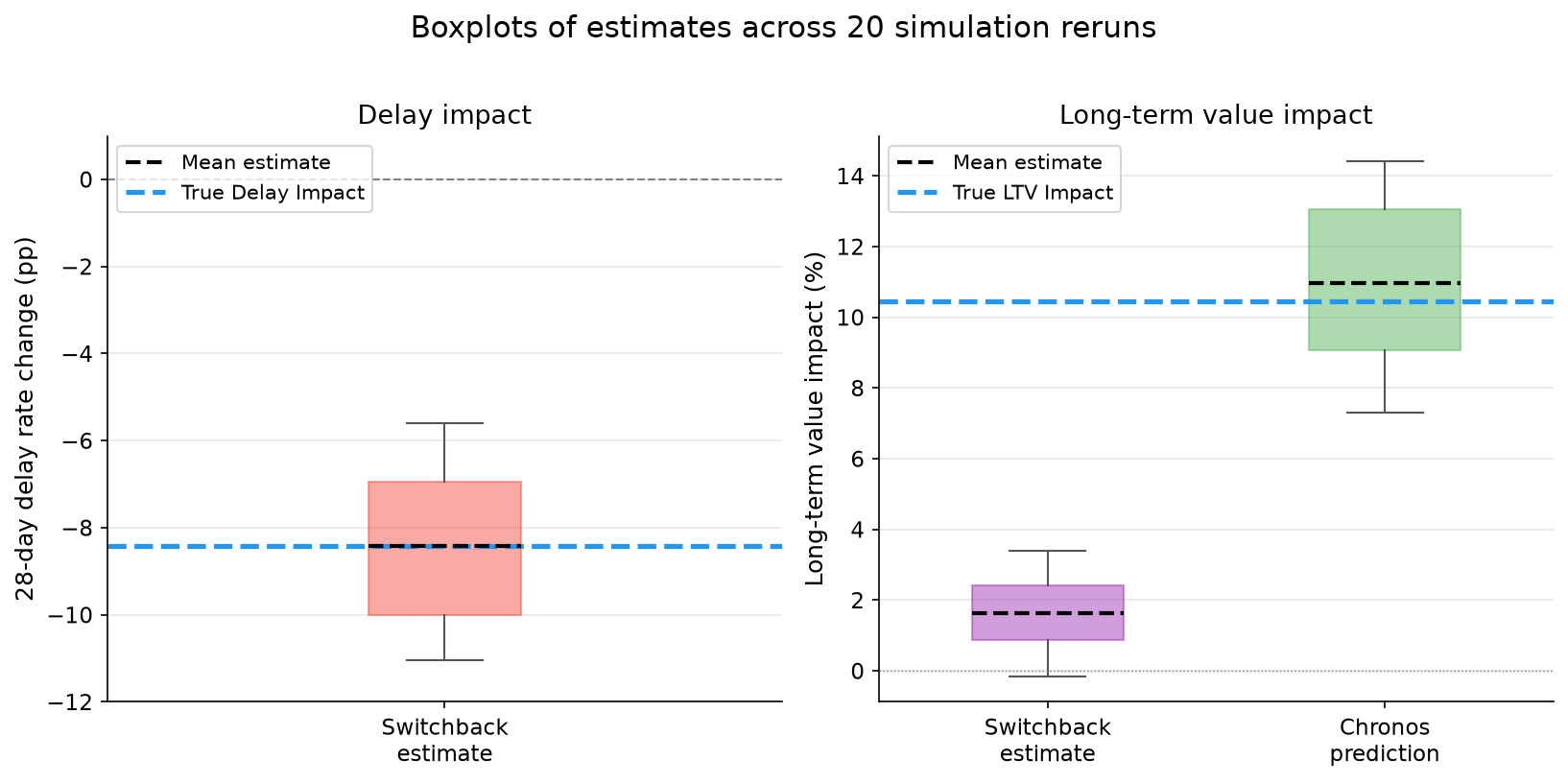}
    \caption{Results across 20 eater-path reruns of a simulated marketplace under increased supply policy change. {\it Left:} Distribution of switchback estimate on 28-day delay exposure impact (mean $-8.42$ pp, std 1.68 pp) against the true delay impact in the first 28 days (dashed line, $-8.42$ pp). {\it Right:} Distribution of switchback eater value impact readout (mean $+1.62$\%, std 0.95\%) and Chronos prediction of long-term impact (mean +10.98\%, std 2.19\%) against the true long-term-value impact (dashed line, $+10.45$\%) }
    \label{fig:boxplots}
\end{figure}

Figure \ref{fig:boxplots} demonstrates the feasibility of our proposed ``switchback $\times$ Chronos'' approach in recovering the total policy effect. The left panel shows that the simulated switchback experiment accurately recovers the short-term equilibrium impact of extra supply on delays. The switchback estimates, centered at \(-8.42\) percentage points, match the true 28-day delay impact shown by the reference line. This confirms that the switchback design captures the full market-level effect on delays, including short-term spillovers through shared courier supply. 
The right panel shows why the switchback alone cannot be further used for long-run policy evaluation. The direct switchback value estimate is centered around only \(+1.62\%\), far below the true long-term value impact of \(+10.45\%\). This gap arises because the 28-day switchback captures only contemporaneous eater responses during the switching window, but not the downstream eater lifecycle effects that accumulate after delay improvements persist. By contrast, the Chronos prediction combines the simulated switchback delay estimate with the Chronos LTV estimate. This combined estimate is centered at \(+10.98\%\), close to the true long-run LTV impact. 

\begin{table}[t]
\centering
\small
\begin{tabular}{p{0.25\linewidth}ccccc}
\toprule
\textbf{Estimator} & \textbf{Mean Est.} & \textbf{Mean True Eff.} & \textbf{Bias} & \textbf{RMSE} & \textbf{Covg.} \\
\midrule

Switchback delay
\((\tau_{\mathrm{DM}})\)
& \(-8.21\) pp
& \(-8.48\) pp
& \(+0.27\) pp
& \(2.51\) pp
& \(96\%\) \\

Switchback $\Delta \mathrm{LTV}$ 
& \(+2.44\%\)
& \(+10.93\%\)
& \(-8.49\%\)
& \(8.80\%\)
& \(6\%\) \\

Chronos $\Delta \mathrm{LTV}$
\((\tau_{\mathrm{DM}} \cdot \beta)\)
 & $+10.71\%$ & $+10.93\%$
  & $-0.23\%$ & $3.71\%$ & $92\%$ \\

\bottomrule
\end{tabular}
\caption{Estimator performance against ground truth across 200 independent market simulations. Bias and RMSE are computed against each simulation's paired ground-truth causal impact; coverage is the empirical fraction of replications in which the estimator's nominal 95\% confidence interval contains the ground truth. 
For Chronos (CBPS-IPW) estimate $\beta=-1.30$\%/pp is estimated once on the observational data and CI is computed with 100 bootstrap resamples. The standard error for the product estimator \eqref{eq:prod_est} is obtained via the delta method, treating the switchback delay and Chronos LTV estimates as independent. Coverage is reported for normal 95\% confidence intervals.}
\label{tab:estimator_performance}
\end{table}

We also run a larger study with 200 \rev{independently regenerated evaluation phase forks}. Every run draws its own eater demand, courier supply, and supply/demand shocks during the forked evaluation phase. This lets us evaluate whether Chronos estimators work reliably across different simulated market conditions, not just in one particular market realization. 
Table \ref{tab:estimator_performance} presents the bias, root mean squared error, and coverage achieved by each estimator across the 200 replications. The switchback estimate on market-level delay exposure change is effectively unbiased and its confidence interval attains 96\% coverage of the true impact.\footnote{\rev{The 200 replications regenerate only the forked evaluation phase; the observational eater history and corresponding Chronos coefficient $\beta=-1.30$\%/pp are held fixed. The eater-clustered bootstrap accounts for eater-level sampling uncertainty from the same eater population in observation phase. The final estimate is a product of switchback estimate and Chronos coefficient. We treat these two components as approximately independent when computing the standard errors as they are estimated from separate observational and evaluation samples and periods. The coverage is evaluated over repeated marketplace realizations in the evaluation phase, not a full end-to-end regeneration.}} In contrast, the switchback estimate on the eater value change only reflects the contemporaneous spend change and is missing roughly three-quarters of the full policy impact: It recovers only +2.44\% against a true impact of +10.93\% and its confidence interval covers the truth in just 6\% of runs. The Chronos prediction corrects the bias almost entirely (bias -0.23\% against +10.93\%; RMSE 3.71\%), reducing the magnitude of the bias by more than thirtyfold relative to the switchback readout.

\section{Discussion}

There is a burgeoning literature on methods that rely on various policy-gradient representations to estimate marginal policy effects in dynamic systems from short-term interventions \citep{farias2022markovian,ghosh2025non,johari2025estimation,lai2026estimating}. Here, we demonstrated promise of this approach in estimating business-relevant long-term metrics in the context of a marketplace for on-demand food delivery. We consider a general model where individual customers operate under Markovian dynamics, but with state variables (such as beliefs and preferences) that remain unobservable to the analyst. Chronos LTV enables us to quantify the long-term cost of service delays from observational data under such general dynamics, thus enabling us to calibrate the marketplace such as to optimize service quality and customer experience. 

In this paper, we used Chronos LTV to estimate the long-term benefits of reliability on a user-by-user level. There are, of course, many other relevant dimensions of reliability we didn't consider. Any market-level change is susceptible to create spillover effects, and correcting for such spillovers may be necessary \citep{munro2025treatment}. Furthermore, here, we only considered the benefits of reliability through the channel of making delays more or less rare; in practice, however, one may need to consider effects of delivery on average and/or estimated delivery times. Finally, we here considered the question of paying for improved reliability in a vacuum, but in fact investing in reliability may come at the expense of outside options such as brand marketing. Practical use of Chronos in a real-world setting will most likely involve applying the method within the context of a broader analytic framework.

\section{Proof of Theorem \ref{theo:pgt}}
\label{sec:proof}

First, under Assumption \ref{assu:markov}, the standard policy-gradient theorem \citep{sutton1999policy} immediately characterizes our target estimand as
\begin{equation}
\label{eq:pgt}
\tau = \frac{1}{T} \sum_{t = 1}^T \EE{O_{it} \p{\EE{\sum_{s = t}^T Y_{is} \cond U_{it}, \, X_{it}, \, D_{it} = 1} 
- \EE{\sum_{s = t}^T Y_{is} \cond U_{it}, \, X_{it}, \, D_{it} = 0}}},
\end{equation}
where $O_{it} = 1\p{\cb{X_{it} \neq \emptyset}}$ is short-hand for whether an order occurred.
Meanwhile, direct manipulation of expectations reveals that
\begin{align*}
\EE{\htau^*_K} &= \frac{1}{T} \sum_{t = 1}^T \EE{O_{it}\p{\frac{D_{it}}{\pi_t(X_{it})} - \frac{1 - D_{it}}{1 - \pi_t(X_{it})}} \, \Gamma_{it}^K} \\
&= \frac{1}{T} \sum_{t = 1}^T \EE{O_{it} \, \EE{\p{\frac{D_{it}}{\pi_t(X_{it})} - \frac{1 - D_{it}}{1 - \pi_t(X_{it})}} \, \Gamma_{it}^K \cond X_{it}}} \\
&= \frac{1}{T} \sum_{t = 1}^T \EE{O_{it} \, \EE{{\frac{D_{it} \, \EE{\Gamma_{it}^K \cond X_{it}, \, D_{it} = 1}}{\pi_t(X_{it})} - \frac{\p{1 - D_{it}}\EE{\Gamma_{it}^K \cond X_{it}, \, D_{it} = 0}}{1 - \pi_t(X_{it})} }\cond X_{it}}} \\
&= \frac{1}{T} \sum_{t = 1}^T \EE{O_{it} \, \EE{{\EE{\Gamma_{it}^K \cond X_{it}, \, D_{it} = 1} - \EE{\Gamma_{it}^K \cond X_{it}, \, D_{it} = 0}} \cond X_{it}}} \\
&= \frac{1}{T} \sum_{t = 1}^T \EE{O_{it}\p{\EE{\Gamma_{it}^K \cond X_{it}, \, D_{it} = 1} - \EE{\Gamma_{it}^K \cond X_{it}, \, D_{it} = 0}}}.
\end{align*}
By Assumption \ref{assu:mix}, the effect of truncating the $\Gamma_{it}^K$ at horizon $K$ can then be verified to decay exponentially in $K$,
\begin{align}
\notag
&\EE{\htau^*_K} =  \frac{1}{T} \sum_{t = 1}^T \EE{O_{it}\p{\EE{\sum_{s = t}^T Y_{is} \cond X_{it}, \, D_{it} = 1} 
- \EE{\sum_{s = t}^T Y_{is} \cond X_{it}, \, D_{it} = 0}}} + R, \\
\label{eq:tau*expr}
&\abs{R} \leq C / (1 - e^{-1/\nu}) \, e^{-K/\nu}.
\end{align}
Finally, by Assumption \ref{assu:unconf}, for $d \in \cb{0, \, 1}$ and all $t$,
\begin{align*}
&\EE{O_{it} \, \EE{\sum_{s = t}^T Y_{is} \cond U_{it}, \, X_{it}, \, D_{it} = d}}
= \EE{O_{it} \, \EE{\EE{\sum_{s = t}^T Y_{is} \cond U_{it}, \, X_{it}, \, D_{it} = d} \! \cond X_{it}}} \\
&\quad\quad= \EE{O_{it} \, \EE{\EE{\sum_{s = t}^T Y_{is} \cond U_{it}, \, X_{it}, \, D_{it} = d} \! \cond X_{it}, \, D_{it} = d}} \\
&\quad\quad = \EE{O_{it} \, \EE{\sum_{s = t}^T Y_{is} \cond X_{it}, \, D_{it} = d}},
\end{align*}
where the first and third equalities are by the chain rule and the second is by conditional independence of delays. Using this fact to compare \eqref{eq:pgt} and \eqref{eq:tau*expr} completes the proof.

\bibliographystyle{plainnat}
\bibliography{references}

\FloatBarrier
\newpage

\begin{appendix}

\section{Simulator Details}
\label{sec:simu_details}
This simulator is provided for illustrative purposes only. It does not reflect actual production systems, policies or operating parameters of any real-world marketplace.
Replication files are available at at \url{https://github.com/chenyuqiu/ltv_of_reliability}.

\subsection{Overview}
This section describes the delivery platform simulator used in Section \ref{sec:simulator}, \ref{sec:exp1} and \ref{sec:exp2}. It is a discrete-time, agent-based process that generates synthetic panel data and captures stylized features in a delivery marketplace: Heterogeneous users with persistent behavioral memory, a multi-sided marketplace with stochastic courier supply, and state variables (such as beliefs and preferences) that remain unobservable to the analyst.

The simulation operates in discrete daily time steps, with each day proceeding through four sequential stages:

\begin{enumerate}
    \item \textbf{New eater arrivals}: A Poisson flow of newly registering eaters is appended to the population.
    \item \textbf{Demand realization}: Each user generates daily orders with order time and characteristics, based on their latent preferences, current delay memory and lifecycle state.
   \item \textbf{Marketplace matching}: An inner loop, discretized into 5-minute intervals, matches realized orders to a stochastic supply of couriers. Any order whose realized completion time exceeds expected delivery time by more than 20 minutes is classified as a delay.
   \item \textbf{Eater state evolution}: At the end of each simulated day, eater-specific state variables (e.g., delay memory, recency, rolling frequency and spending) are updated based on the day's realized marketplace outcomes. Eaters then transition between lifecycle states according to a Markov chain conditioned on these newly updated variables.
\end{enumerate}

\subsection{Eater Generative Model}

Every user is endowed with a 7-dimensional latent parameter vector drawn once from a joint log-normal distribution that remains invariant over the simulation:
\begin{equation}
    \mathbf{\theta}_i \sim \text{LogNormal}(\boldsymbol{\mu}_{\text{pop}}, \boldsymbol{\Sigma}_{\text{pop}})
\end{equation}
where $\boldsymbol{\Sigma}_{\text{pop}} = \text{diag}(\boldsymbol{\sigma}) \cdot \mathbf{C} \cdot \text{diag}(\boldsymbol{\sigma})$, and $\mathbf{C}$ is a symmetric $7 \times 7$ correlation matrix capturing economically motivated co-dependencies across latent traits.
Table \ref{tab:latent_parameters} describes these latent parameters.
Furthermore, each simulation day, each user's behavioral paths evolve through a vector of mutable state variables which track user lifecycle stage, historical experiences, and rolling spendings; see Table \ref{tab:mutable_state}.

\begin{table}[p]
\centering
\small
\begin{tabular}{lp{5.2cm}p{7.5cm}}
\toprule
Symbol & Parameter & Key Correlations \\
\midrule
$\beta_y$ & Baseline order rate & \textbf{Loyalty Cluster:} Positively correlated with baseline spending ($\beta_{\text{spend}}$) and tipping ($\beta_{\text{tip}}$); negatively correlated with delay sensitivity ($\beta_d$). \\
\addlinespace
$\beta_d$ & Delay sensitivity & \textbf{Habitual Resilience:} Negatively correlated with baseline demand ($\beta_y$) and baseline spending ($\beta_{\text{spend}}$).\\
\addlinespace
$\beta_{\text{spend}}$ & Baseline spending & \textbf{Power-User Cluster:} High-frequency users possess larger shopping baskets, but tend to exhibit a negative correlation with explorativeness ($\sigma_y$). \\
\addlinespace
$\beta_{\text{tip}}$ & Tip amount & \textbf{Friction Compensation:} Strongly positively correlated with delivery distance ($\mu_{\text{dist}}$) and kitchen prep times ($\mu_{\text{prep}}$) to attract couriers. \\
\addlinespace
$\sigma_y$ & Explorativeness & \textbf{Variety Seeking:} Positively correlated with baseline rate ($\beta_y$); frequent buyers diversify choices. \\
\addlinespace
$\mu_{\text{dist}}$ & Delivery distance & \textbf{Restaurant Tiering:} Positively correlated with prep times ($\mu_{\text{prep}}$); geographically distant options heavily cluster with premium, slower merchants. \\
\addlinespace
$\mu_{\text{prep}}$ & Restaurant prep time & \textbf{Supply-Side Friction:} Long preparation times structurally co-vary with high distance ($\mu_{\text{dist}}$) and courier tipping premiums ($\beta_{\text{tip}}$). \\
\bottomrule
\end{tabular}
\caption{Latent User Parameters and Behavioral Correlation Structure}
\label{tab:latent_parameters}
\end{table}

\begin{table}[p]
\centering
\small
\begin{tabular}{llp{9.5cm}}
\toprule
Symbol & State Variable & Description \\
\midrule
$L_{it}$ & Lifecycle Category & Discrete state index: $L_{it} \in \{\text{New}, \text{Casual}, \text{Power}, \text{At-Risk}, \text{Churned}\}$ \\
\addlinespace
$D_{it}$ & Delay Memory & Continuous impact metric: $D_{it} \in (0, 1]$, where 1 represents unsuppressed baseline utility \\
\addlinespace
$R_{it}$ & Recency Count & Elapsed calendar days since the eater's most recent order \\
\addlinespace
$F_{it}^{(30)}$ & Rolling Frequency & Count of cumulative orders placed within the trailing 30-day window. \\
\addlinespace
$S_{it}^{(30)}$ & Rolling Spending & Cumulative value transacted within the trailing 30-day window. \\
\addlinespace
$S_{it}^{(\text{total})}$ & Lifetime Spending & Total cumulative value transacted by the eater since start \\
\addlinespace
$N_{it}^{(\text{total})}$ & Lifetime Orders & Total cumulative order count completed by the eater since start. \\
\bottomrule
\end{tabular}
\caption{Mutable Per-User State Variables}
\label{tab:mutable_state}
\end{table}

The influx of new eaters is modeled as a Poisson process:
\begin{equation}
    n_{\text{new}, t} \sim \text{Poisson}(\lambda_{\text{growth}} \cdot n_{\text{active}, t})
\end{equation}
This structural growth rate is designed to balance out the long-run per-capita churn rate. Given a baseline operating scale of $n_{\text{active}} \approx 100,000$ users, this process infuses approximately $100$ new consumer arrivals per day. 
Newly arrived eaters enter the marketplace with an initial state, uniformly at the \textit{New} lifecycle state ($L_{i0} = \text{New}$), no delay memories ($D_{i0} = 1$), and zeroed recency profiles ($R_{i0} = 0$).

\subsection{Eater Behavior Dynamics}

\paragraph{Delay Memory Dynamics} Delay memory $D_{it}$ models the behavioral suppression of an eater's future ordering propensity following a delivery delay. The state updates daily according to a geometric decay rule:
\begin{equation}
    \label{eq:delay_memory_update}
    \log D_{i, t+1} = e^{-\lambda_{\text{mem}}} \log D_{it} - \beta_{d,i} \cdot \mathbf{1}[\text{delayed today}]
\end{equation}
where $\lambda_{\text{mem}} = 0.06$, implying an experimental half-life of 11.6 days. While memory recovery decay is applied universally across all agents each period, additional delay incurs further drops in the order propensity differently driven by idiosyncratic latent sensitivity $\beta_{d,i}$.

For a typical \textit{New} or \textit{At-Risk} eater ($\beta_{d,i} \approx 0.37$), delay memory $D_{it}$ degrades sequentially from $1.00 \to 0.69 \to 0.49 \to 0.35$ across zero, one, two, and three delays, respectively. Memory recovery naturally occurs with no delay or over extended periods of inactivity: a churned eater spending 90 days away from the platform recovers $D_{it}$ towards 1 upon re-entry, effectively resetting to a neutral behavioral baseline.

\begin{figure}[t]
    \centering
    \includegraphics[width=1\linewidth]{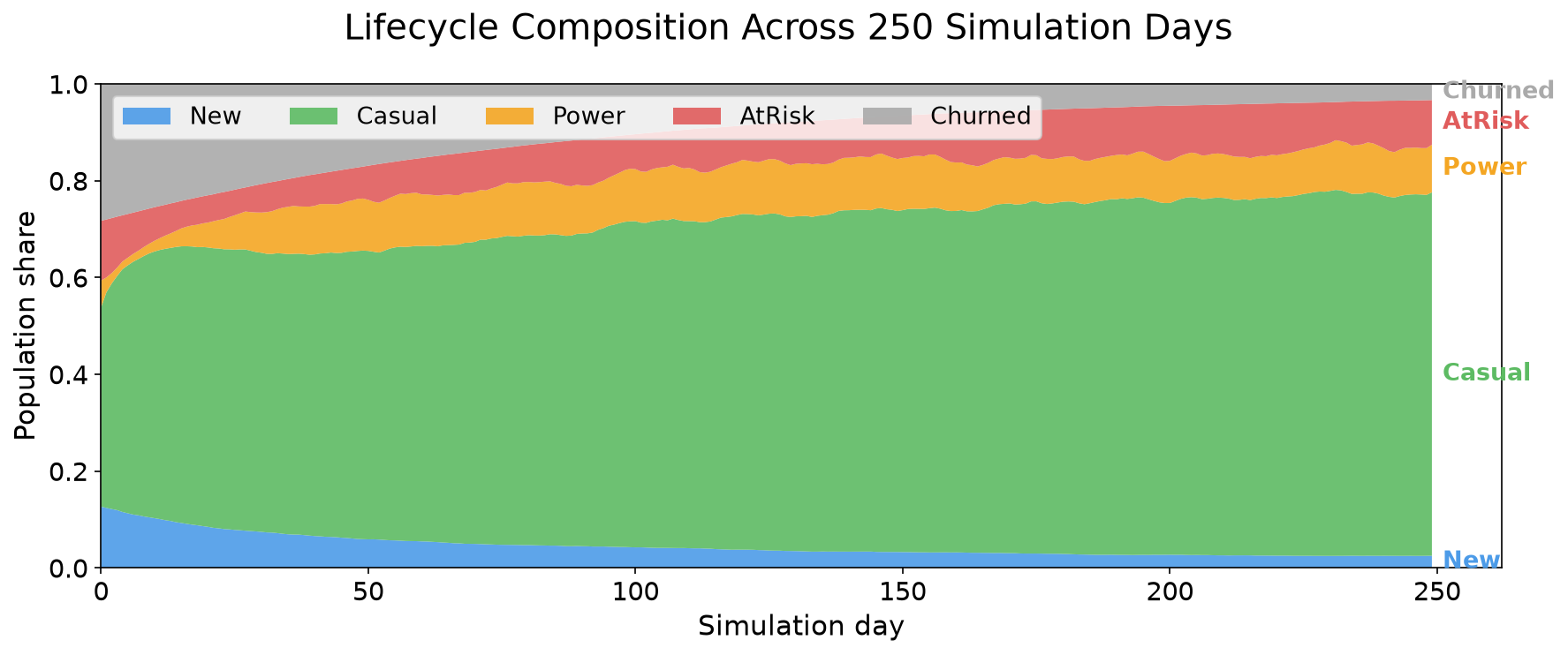}
    \caption{Lifecycle composition of the simulated population across 250 simulation days.}
    \label{fig:lifecycle_composition}
\end{figure}

\paragraph{Lifecycle Transitions} The eater population distribution moves dynamically across five distinct customer segments over time that are not observable to the analyst: \rev{New (newly arrived or reactivated), Casual (moderate ordering activity), Power (high order activity), At-Risk (low ordering activity), Churned (0 ordering activity but may return after some time).} Figure~\ref{fig:lifecycle_composition} tracks the aggregate lifecycle composition of the simulated population across 250 baseline simulation days, illustrating the steady state achieved under these dynamics. 
Eater movements across these five states are modeled as a discrete-time Markov chain. The daily conditional transition probabilities are parameterized via logistic functions of observable behavioral history, operating along the following diagram:
\begin{align*}
    \text{New} &\rightarrow \text{Casual} \rightarrow \text{Power} \\
    \bigg\downarrow &\ \ \ \swarrow \ \ \ \ \swarrow \\
    \text{At-Risk} &\qquad \xrightarrow{\text{\scriptsize Recovery}} \text{Casual} \\
    \bigg\downarrow & \\
    \text{Churned} &\qquad \xrightarrow {\text{\scriptsize Reactivation}} \text{New}
\end{align*}
Table \ref{tab:transition_reference} describes these transitions in more detail.

\begin{table}[t]
\centering
\small
\begin{tabular}{llp{9.5cm}}
\toprule
From State & To State & Covariates that Affect Probability and Directions \\
\midrule
\textbf{New} & Casual & \textbf{(+)} Cumulative lifetime orders, ordering today \\
& & \textbf{(-)} Accumulated log delay memory \\
\addlinespace
\textbf{New} & At-Risk & \textbf{(+)} Recency days, log delay memory \\
& & \textbf{(-)} Lifetime orders, ordering today \\
\addlinespace
\textbf{Casual} & Power & \textbf{(+)} Trailing frequency, rolling spending, lifetime orders, ordering today \\
& & \textbf{(-)} Recency days, log delay memory \\
\addlinespace
\textbf{Casual} & At-Risk & \textbf{(+)} Recency days, log delay memory \\
& & \textbf{(-)} Lifetime orders, ordering today \\
\addlinespace
\textbf{Power} & At-Risk & \textbf{(+)} Recency days, log delay memory \\
& & \textbf{(-)} Trailing frequency, rolling spending \\
\addlinespace
\textbf{Power} & Casual & \textbf{(+)} Log delay memory, recency days \\
& & \textbf{(-)} Trailing frequency, rolling spending \\
\addlinespace
\textbf{At-Risk} & Casual & \textbf{(+)} Log delay memory, trailing frequency, rolling spending, lifetime orders, ordering today \\
& & \textbf{(-)} Recency days \\
\addlinespace
\textbf{At-Risk} & Churned & \textbf{(+)} Recency days, log delay memory \\
& & \textbf{(-)} Trailing frequency, rolling spending, lifetime orders, ordering today \\
\bottomrule
\end{tabular}
\caption{Summary of Lifecycle Transition Model}
\label{tab:transition_reference}
\end{table}

\subsection{Multi-Sided Delivery Market}

\paragraph{Demand Model} At the start of each simulation day $t$, number of orders placed by eater $i$ on tha day are drawn from a non-homogeneous Poisson process:
\begin{equation}
    n_i(t) \sim \text{Poisson}(\lambda_i(t))
\end{equation}
The underlying intensity parameter $\lambda_i(t)$ is modeled log-linearly to incorporate eater-specific traits, delay memory and macro shocks:
\begin{equation}
 \lambda_i(t) =  \beta_{y,i} \cdot D_i(t) \cdot m(L_{it}) \cdot \kappa_t 
\end{equation}
where each component is structured as follows:
\begin{itemize}
    \item $m(L_{it}) \in \{0.3, 0.6, 1.0, 0.2, 0.0\}$ represents the lifecycle state multiplier assigned to the categorical vector $L_{it} \in \{\text{New}, \text{Casual}, \text{Power}, \text{At-Risk}, \text{Churned}\}$.
    \item $\kappa_t \sim \text{LogNormal}(0, \sigma_{\text{demand}}^2)$ with $\sigma_{\text{demand}}$ captures a uniform daily macroeconomic demand shock (e.g., severe weather events, holidays). In expectation it introduces roughly $10\%$ day-to-day variation in aggregate platform demand.
\end{itemize}

For each realized order, the order timestamp is modeled using a complex multimodal normal-uniform mixture distribution tracking daily fulfillment patterns:
\begin{align}
    \text{order\_time}_k &\sim 0.34 \cdot \mathcal{N}(720, 45^2) + 0.44 \cdot \mathcal{N}(1140, 60^2) + 0.10 \cdot \mathcal{N}(1350, 30^2) \nonumber \\
    &\quad + 0.08 \cdot \text{Uniform}(420, 1380) + 0.04 \cdot \text{Uniform}(0, 360)
\end{align}
This mixture cleanly captures peak lunch (12:00), peak dinner (19:00), late-night demand (22:30), daytime background volume (07:00--23:00), and overnight orders (00:00--06:00).

Eater-specific order attributes, including delivery distance, restaurant prep time, basket size, and tip, are drawn from conditional log-normal distributions per eater's latent traits. The idiosyncratic explorativeness parameter $\sigma_{y,i}$ governs the variance of all four order-level attributes simultaneously. Highly habitual eaters exhibit low variance (producing similar baskets and distances), while exploratory eaters generate substantial variance across trip dimensions, merchant types, and spending sizes.

\paragraph{Marketplace} The daily marketplace executes an inner loop discretized into 5-minute intervals, yielding 288 discrete clearing slots per 24-hour cycle. At the start of each daily cycle, the platform initializes the matching parameters:
\begin{enumerate}
    \item \textbf{Supply Shock Realization:} A daily system supply factor is drawn from $\xi_t \sim \text{LogNormal}(0, \sigma_{\text{supply}}^2)$. This specifies rare but highly severe fulfillment disruptions.
    \item \textbf{Fulfillment Noises:} Independent delivery frictions are drawn for travel velocity / congestion ($\varepsilon_k^{\text{travel}} \sim \text{LogNormal}(0, \sigma_{\text{travel}}^2)$) and kitchen preparation time ($\varepsilon_k^{\text{kitchen}} \sim \text{LogNormal}(0, \sigma_{\text{kitchen}}^2)$).
    \item \textbf{Queue Serialization:} All incoming orders are indexed in a First-In, First-Out (FIFO) sequence based on queue entry time, which is based on order placement time + food preparation time. In AB experiment validation, this is just food ready time for control orders, but for treated orders: \begin{equation}
    \text{Food Prep Time} = \max(\text{Kitchen Time} - 5\text{ mins}, 0)
\end{equation}
thus they get dispatched into the FIFO queue earlier.
\end{enumerate}

\paragraph{Courier Arrivals} The arrival of new couriers into the marketplace pool within any given 5-minute slot is governed by a Negative Binomial process:
\begin{equation}
    n_{\text{couriers}, t} \sim \text{NegBin}\left(r, \frac{r}{r + \lambda_{\text{slot}}}\right)
\end{equation}
where $r$ is the structural overdispersion parameter. The time-varying intensity parameter $\lambda_{\text{slot}}$ is defined as:
\begin{equation}
    \lambda_{\text{slot}} = C[\text{dow}, \text{hour}] \cdot \xi_t \cdot \frac{n_{\text{eaters}}}{700} \cdot \text{multi}_{\text{sb}}
\end{equation}
where $C$ is a $(7 \times 24)$ baseline arrival matrix, $\xi_t$ is realized supply shock. The scaling factor $n_{\text{eaters}}/700$ dynamically scales courier density alongside eater population growth to maintain stable systemic utilization. $\text{multi}_{\text{sb}}$ is used in switchback simulation to implement extra courier policy: for treated-block hours, it is set to 1.25x and control-block hours to 1.0x.

\paragraph{Matching Mechanism} During each 5-min slot, the platform processes the persistent courier pool sequentially using a FIFO assignment rule. Each active courier $j$ evaluates up to the top $K=4$ pending orders in the queue, accepting the first order that satisfies a logistic probability threshold:
\begin{equation}
    p_{\text{accept}}(j, k) = \sigma\left(\alpha_{c,j} - \gamma_{\text{dist}} \cdot d_k + \gamma_{\text{tip}} \cdot \text{tip}_k - \mathbf{1}[\text{peak}] \cdot \gamma_{\text{peak}}\right)
\end{equation}
where $\alpha_{c,j}$ is a latent tolerance parameter assigned to each courier upon market entry. The acceptance probability is positively affected by the tip and negatively affected by the distance. The peak penalty $\mathbf{1}[\text{peak}]$ applies during lunch (11:00--13:00) and dinner (18:00--21:00) intervals, modeling heightened courier selectivity during peak hours. 

Matched couriers immediately exit the active pool to complete delivery operations and do not rejoin the market. Unmatched couriers persist in the pool across slots until encountering a maximum wait timeout of 30 minutes. Conversely, pending orders expire and are canceled if they remain unassigned in the queue for more than 80 minutes. This unfulfilled path yields zero spending value and is excluded from delay classification, maintaining an average baseline cancellation rate below $5\%$.

\paragraph{Delay outcome} An order $k$ is formally classified as delayed if its total realized fulfillment duration ($\text{actual\_dur}_k$) exceeds the initially promised delivery time ($\text{ETA}_k$) by more than a fixed buffer:
\begin{equation}
    \text{delayed}_k = \mathbf{1}\left[\text{actual\_dur}_k > \text{ETA}_k + \delta_{\text{buf}}\right]
\end{equation}
where the classification margin is fixed at $\delta_{\text{buf}} = 20\text{ minutes}$. The eater-facing $\text{ETA}_k$ is quoted at checkout using baseline nominal parameters across both arms of the simulation:
\begin{equation}
    \text{ETA}_k = \tau_{\text{dist}} \cdot d_k + p_k + E_{\text{match}} + \eta_{\text{buffer}}
\end{equation}
with $d_k$ and $p_k$ as the distance and prep time generated with the order, $\tau_{\text{dist}}$ is base travel rate and platform safety buffer $\eta_{\text{buffer}} = 5\text{ min}$. $E_{\text{match}}$ is the expected marketplace queue waiting time, updated daily via an exponentially weighted moving average with smoothing factor.

The total realized fulfillment duration can be broken down into two sequential stages: restaurant pickup and delivery transit.
\begin{equation}
    \text{actual\_dur}_k = \max\underbrace{\left(p_k \cdot \varepsilon_k^{\text{kitchen}}, \ p_k^{\text{eff}} + m_k\right)}_{\text{Order to Restaurant Pickup}} + \underbrace{\left(\tau_{\text{dist}} \cdot d_k \cdot \varepsilon_k^{\text{travel}}\right)}_{\text{Pickup to Delivery }}
\end{equation}
where $\varepsilon_k^{\text{travel}}, \varepsilon_k^{\text{kitchen}}$ represent independent travel and preparation noise multipliers, and $m_k$ is the realized FIFO match time.

\begin{table}[t]
\centering
\small
\begin{tabular}{p{0.24\linewidth} p{0.68\linewidth}}
\toprule
\textbf{Feature} & \textbf{Description} \\
\midrule

\texttt{delay\_rate\_30d} 
& Fraction of the eater’s orders that experienced delivery delays during the past 30 simulated days. \\

\texttt{delay\_rate\_7d} 
& Fraction of the eater’s orders that experienced delivery delays during the past 7 simulated days. \\

\texttt{F30} 
& Total number of orders placed by the eaters during the past 30 simulated days. \\

\texttt{S30} 
& Total spending by the eater during the past 30 simulated days. \\

\texttt{R} 
& Eater recency, measured as the number of simulated days since the eater’s last order. \\

\texttt{n\_orders\_lifetime} 
& Cumulative number of lifetime orders placed by the eater. \\

\texttt{distance} 
& Delivery distance associated with the current order. \\

\texttt{prep\_time} 
& Restaurant food preparation time associated with the current order. \\

\texttt{tip} 
& Tip amount associated with the current order. \\

\texttt{day\_of\_week} 
& Day of week capturing weekly demand seasonality. \\

\texttt{hour\_of\_day} 
& Hour of day capturing intra-day demand patterns. \\

\bottomrule
\end{tabular}
\caption{Order- and Eater-level Features Used in Chronos Estimation}
\label{tab:ltv_features}
\end{table}

The $\max(\cdot)$ operator encodes the proactive dispatch we implement into the simulation for AB experiment validation. Under baseline food-ready dispatch, the control orders only enter the queue after the cooking finishes ($p_k^{\text{eff}} = p_k $), forcing preparation and courier assignment to happen sequentially. Under the proactive dispatch treatment arm, the platform attempts to run these channels concurrently by dispatching the courier 5 minutes before food preparation finishes:
\begin{equation}
    p_k^{\text{eff}} = \max(p_k - 5, 0)
\end{equation}
This structural concurrency shortens $\text{actual\_dur}_k$ by up to 5 minutes without altering the consumer's promised $\text{ETA}_k$, directly mitigating delays.

It is important to highlight the key structural confounding mechanisms operating within the simulation:
\begin{enumerate}
    \item \textbf{Order-Characteristic Selection:} The logistic courier acceptance model specifies that a courier's willingness to accept an assignment decays with delivery distance ($d_k$) and increases with the tipping premium ($\text{tip}_k$). As a result, longer-distance or lower-tip orders experience longer queue wait times and face a structurally higher probability of delay. 
    \item \textbf{Day-Level Supply Shock Correlation:} Systemic platform-wide capacity varies daily via the aggregate supply factor $\xi_t$. On low-supply days, the entire order pool faces heightened delay rates uniformly, regardless of individual eater profiles. Day-of-week effects and aggregate rolling history delay rates generate strong temporal correlations. 
\end{enumerate}
These paths correlate directly with observable eater characteristics, including historical rolling delay metrics, past order frequency, trailing spending, and recency, and thus require a confounding adjustment as provided by Theorem \ref{theo:pgt}. The observed features available for estimation are given in Table \ref{tab:ltv_features}.

\section{\rev{Additional Simulation with Unobserved Confounder}}

\label{sec:unobserved-confounding}

\rev{In our delivery delay application, sequential unconfoundedness is plausible because as the platform we observe a rich set of pre-delay information, such as order and trip characteristics as well as historical data of eaters' prior experiences. Conditioning on this information should be sufficient to make delay exposure as-if random. In this section, we modify the baseline simulator to examine how Chronos behaves when the assumption is only approximately satisfied or strongly violated.}

\rev{To this end, we consider adding (random and unobserved) kitchen operational problems to the model. Such kitchen issues simulatenously slow down cooking (thus causing increased delays) and lead to food quality issues (thus reducing future order probability)---thereby violating sequential unconfoundedness. Specifically, we introduce a latent shock,
$U_{it} \sim \operatorname{Bernoulli}(0.20)$,
drawn independently across eaters and days.
The shock affects the delivery time by increasing realized kitchen time via the relation $\varepsilon_k^{\text{kitchen}}e^{0.5U_{it}}$.
Meanwhile, the shock affects future demand directly (i.e., through the food quality channel) as follows. Let \smash{$H_{it}^{(M)}$} denote the number
of times an eater was exposed to kitchen quality issues over the preceding $M$ days. Daily ordering probability is then modified by
\[
\lambda_{it}
=
\lambda_{it}^{\mathrm{base}}
\exp\!\left(-\gamma H_{it}^{(M)}\right).
\]
One exposure therefore reduces ordering propensity by a multiplicative factor $\exp(-\gamma)$; when $\gamma = 0.3$, this is about 26\% lower.}

\begin{table}[t]
\centering
\small
\begin{tabular}{lrrrrrr}
\toprule
Scenario
& Chronos $\bar\beta$
& A/B $\Delta$delay
& Chronos $\Delta$LTV
& A/B $\Delta$LTV
& Calibr.\\
& (\% per pp)
& (pp) & (\%) & (\%) &  \\
\midrule
$\gamma=0$
  & $-1.344$ & $-3.445$ & $+4.630$ & $+4.875$ & $0.95\times$ \\
$\gamma=0.30,\ M=7\mathrm{d}$
  & $-1.242$ & $-3.595$ & $+4.467$ & $+4.464$ & $1.00\times$ \\
$\gamma=0.30,\ M=14\mathrm{d}$
  & $-1.234$ & $-2.801$ & $+3.457$ & $+2.688$ & $1.29\times$ \\
\bottomrule
\end{tabular}
\caption{\rev{Performance comparison with an unobserved eater-day quality shock $U_{it}$
using Chronos estimate with $K=70$ days. When $U_{it}=1$, actual kitchen time is multiplied by $\exp(0.5)$ thus increasing delay probability. If the eater
has a matched order, the exposure additionally multiplies daily
ordering propensity by $\exp(-\gamma)$ over the following $M$ days. When $\gamma=0$ there is no future-demand effect. Chronos estimates are
means of 100 eater-cluster bootstrap refits in each scenario's simulated market. The delay-rate change and observed $\Delta$LTV is from each scenario's dispatch AB experiment.}}
\label{tab:sensitivity_k70}
\end{table}

\rev{Table \ref{tab:sensitivity_k70} shows the Chronos prediction results under three scenarios of the added eater-day shock.
When $\gamma=0$, $U_{it}$ affects delay probability but has no direct effect
on future ordering. In this case, the Chronos policy prediction remains close
to the randomized A/B benchmark, with calibration of $0.95\times$. With $\gamma=0.30$ and a 7-day persistence window (i.e., $M = 7)$, the same unobserved shock also affects future demand, but the Chronos prediction is still robust to this slight violation. Extending the future-demand effect to 14 days yields an clear overestimation with calibration of
$1.29\times$.
These results illustrate a possible pattern under violations of
sequential unconfoundedness: As an omitted confounder becomes more strongly
associated with both delay exposure and future outcomes, the
Chronos policy prediction can move further from the benchmark.}

\end{appendix}

\end{document}